\documentclass[aps,pra,twocolumn]{revtex4}
\usepackage{amsmath}
\usepackage{amssymb}
\usepackage{mathrsfs}
\usepackage{epsfig}
\usepackage{color}
\usepackage[bookmarks]{hyperref}
\usepackage{graphicx}
\usepackage{bm}
\usepackage{times}
\usepackage{soul}%%%
\usepackage{caption}
\usepackage{subcaption}

\newcommand*{\bra}[1]{\ensuremath{\langle{#1}\rvert}}
\newcommand*{\ket}[1]{\ensuremath{\lvert{#1}\rangle}}

\DeclareMathOperator{\erf}{erf}

\begin{document}
\title{Optimal estimation of matter--field coupling strength in the dipole approximation}

\author{J\'ozsef Zsolt Bern\'ad}
\email{zsolt.bernad@um.edu.mt}
\affiliation{Department of Physics, University of Malta, Msida MSD 2080, Malta}
\author{Claudio Sanavio}
\email{claudio.sanavio@um.edu.mt}
\affiliation{Department of Physics, University of Malta, Msida MSD 2080, Malta}
\author{Andr\'e Xuereb}
\affiliation{Department of Physics, University of Malta, Msida MSD 2080, Malta}

\date{\today}

\begin{abstract}
This paper is devoted to the study of Bayesian-inference approach in the context of estimating the dipole coupling strength in matter--field interactions. In particular, we consider the simplest model 
of a two-level system interacting with a single-mode of the radiation field. Our estimation strategy is based on the emerging state of the two-level system, whereas we determine both the 
minimum mean-square error and maximum likelihood estimators for uniform and Gaussian prior probability density functions.  In the case of the maximum likelihood estimator, we develop a mathematical method 
which extends the already existing approaches to the variational problem of the average cost function. We demonstrate that long interaction times, large initial mean photon numbers, and non-zero detuning between 
two-level system transition and the frequency of the electromagnetic field mode have a \emph{deleterious} effect on the optimality of the estimation scenario. 
We also present several cases where the estimation process is inconclusive, despite many ideal conditions being met.
\end{abstract}

\maketitle

\section{Introduction}

Measurements of physical systems with complete description can be anticipated and the problem of prediction is the forward problem. The inverse problem consists of the estimation of
physical parameters using the available measurement data~\cite{Kaipio}. In quantum mechanics we seek to estimate the physical parameters governing the evolution of a density matrix from measurements made on part of 
the system. Against this background, the problem of optimal measurements in quantum systems has been a major focus since the beginning of quantum estimation theory~\cite{Helstrom1976, Holevo}. The criterion of 
optimality is defined through the cost experienced upon making errors in the estimates. This measure is formulated by means of the so-called cost function of the estimates and the true values of the parameters. 
The optimum strategy attempts to find that positive-operator valued measure (POVM) which minimizes the average cost functional calculated with the help of the cost function, the density matrix, and the POVM. 
The rigorous mathematical meaning of the average cost functional and conditions under which solutions of the optimization problem exist was thoroughly investigated by Holevo~\cite{Holevo2}. In this context, a 
particularly convenient approach is the Bayesian-inference method, where one assumes that the true values of the physical parameters are random variables with a given prior probability density 
function (p.d.f.)~\cite{Helstrom69}.

In this paper, we continue our investigations of the Bayesian-inference approach with a focus on one-parameter estimation scenarios in order to gain a better insight into the properties of the 
estimators~\cite{Zsolt}. We shall consider the problem of estimating the dipole coupling of matter--field interactions~\cite{Haroche}. Due to the widespread applications of these interactions in, e.g., 
quantum communication~\cite{repeater}, a precise determination of the dipole coupling has increasing technological, as well as fundamental, relevance. While quantum electrodynamics gives a straightforward 
recipe for calculating this matter--field coupling~\cite{Wolfgang}, experimental limitations on precision inherently introduce probabilistic variations in this parameter. Typically, the dipole coupling
varies along the trajectory of the moving atom due to the mode structure in the cavity \cite{Haroche1996,Gleyzes} or in case of trapped atoms due to the temperature induced position probability distribution 
\cite{Isenhower}. The experiments usually determine an effective dipole coupling by integrating over the variations of the coupling strength. Another way to gain some 
knowledge is to perform measurements on the physical system and obtain data, from which the value of the effective dipole coupling can be inferred. In the context of the Bayesian-inference approach
one may even obtain the optimum estimators. Here in this paper, we determine not only the minimum mean--square error estimator for a Gaussian p.d.f.~\cite{Zsolt}, but we  consider also a uniform prior p.d.f. These  
type of a priori p.d.f. are going to be applied also in the determination of optimum maximum likelihood estimator. The method presented in Ref.~\cite{Zsolt} and elaborated upon here should be distinguished from the quantum Fisher 
information approach~\cite{Demkowicz}, which has also been successfully applied to systems with matter--field interactions~\cite{Brunelli1, Brunelli2}.

In our model, two-level systems (TLSs) transit through a cavity supporting a single-mode of the radiation field and are then measured. We trace out the single-mode radiation field and concentrate on the 
resulting density matrix, subject to the quantum estimation procedure. Spontaneous decay of the TLS is also taken into account. In the case of the minimum mean--square error estimator we invoke the method 
applied in our previous work~\cite{Zsolt}, demonstrating that the resulting optimal detection strategy can be related to implementable measurement setups in experimentally relevant situations. The problem 
of determining the maximum likelihood estimator is centered around the resolution of identity and integration with respect to an operator valued measure, see Dobrakov's integral in Ref.~\cite{Dobrakov}. 
Due to our motivations being rooted in physics we choose to avoid generalized theories of the Lebesgue integral~\cite{Yosida}, instead making a simple ansatz for the POVM with the help of square-integrable 
functions. This construction allows us to determine the maximum likelihood estimators for both the uniform and Gaussian prior p.d.f.\ and offers a new mathematical tool
for the maximum likelihood estimation strategy. Whereas estimating the phases of states, displacement parameters, wave vectors, 
and coherent signal amplitudes involves solving the equations for the optimum strategy involving the risk operator~\cite{Holevo2}, here we focus directly
on the extrema of the average cost function, which determine the optimum POVMs. We will present numerical calculations of the average cost functions, the 
average estimates and lower bounds of the mean--squared error of the obtained biased estimators.

This paper is organized as follows. In Sec.~\ref{I} we discuss the model and determine the state of the TLS following its interaction with the single-mode radiation field. Spontaneous decay of the TLS is also 
considered. In Sec.~\ref{II} we recapitulate some basic facts about quantum estimation theory and introduce the formalism used throughout the whole manuscript. We then address the problem of determining the 
minimum mean--square error estimator in Sec.~\ref{III}. In Sec.~\ref{IV} maximum likelihood estimators are discussed. Finally, we discuss our work and draw our conclusions in Sec.~\ref{V}. 

\section{Model}
\label{I}

\begin{figure}
 \includegraphics[width=.39\textwidth]{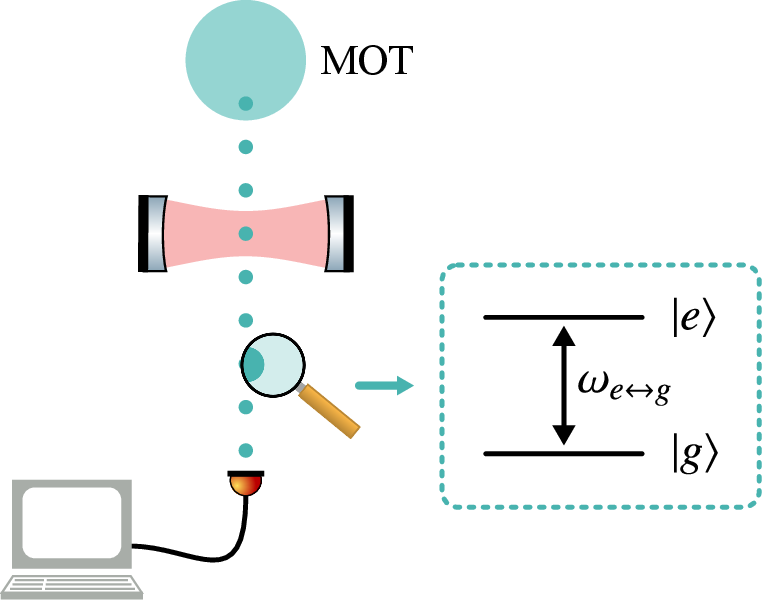}
\caption{Schematic representation of a quantum estimation scenario based on cavity QED. The atoms (grey dots) implementing the two-level systems are captured from a background gas by a 
magneto--optical trap and loaded into an optical conveyor belt~\cite{Meschede2}. The atoms move with the help of the conveyor belt into and out of the cavity and towards a detector. 
The transition frequency of the atom is $\omega_{e \leftrightarrow g}$. Further details about the scheme are in the text.}
   \label{setup}
\end{figure}

In this section we discuss a cavity QED model consisting a TLS interacting with a single-mode electromagnetic cavity. The TLS, generally implemented as a flying atom, 
is injected into the cavity and emerges from the cavity and is detected after interacting with the electromagnetic field. The setup, illustrated in Fig.~\ref{setup}, is one of the 
best suited for our estimation procedure, because it is under exquisite experimental control~\cite{Haroche, Meschede, Meschede2}, and because it allows repeated measurements to be made 
using several TLS interacting sequentially with the field. In fact, this is a very important point in estimation scenarios because the use of $N$ independent and identical systems reduces 
the lower bound of the estimation accuracy by a factor of $N^{-1}$~\cite{Helstrom1976}. Therefore, it is assumed that before each TLS enters the cavity, the single-mode field is always reset to the 
same initial state. The state of each TLS entering the cavity is also assumed to be the same. In practice, the controlled motion of an atom into and out of the cavity may be realized using an optical 
conveyor belt~\cite{Meschede2}, i.e., a moving dipole trap, into which atoms are loaded from a magneto--optical trap. In our discussion, we present the solution to this elementary model and determine the 
state of the atom by tracing out the state of the electromagnetic field. The optimal estimator for the matter--field coupling will be subsequently determined for each presented estimation scenario.

Let us consider a TLS with ground state $\ket{g}$ and excited state $\ket{e}$. Cavity leakage and spontaneous decay of the TLS are present; nonetheless, it is assumed in most parts of the presented work
that the coupling strength of the matter--field interaction is much larger than the damping rate of the two decoherence
sources. Therefore, the joint TLS--field state during the matter--field interaction time can effectively be described by a purely unitary evolution. In the dipole and 
rotating--wave approximations, the Hamiltonian in the time-independent interaction picture reads~\cite{JaynesCumming, Paul} ($\hbar=1$):
\begin{equation}\label{HamJC}
\hat{H}=\tfrac{\Delta}{2}\hat{\sigma}_z + g( \hat{a} \hat{\sigma}_+ + \hat{a}^\dagger \hat{\sigma}_-),
\end{equation}
where $\hat{\sigma}_z=\ket{e} \bra{e}-\ket{g} \bra{g}$, and $\hat{\sigma}_+=\ket{e} \bra{g}$ is the raising and $\hat{\sigma}_-=\ket{g}\bra{e}$ the lowering operator. 
$\hat{a}$ and $\hat{a}^\dagger$ are the annihilation and creation operators of the field mode. $\Delta=\omega_{e \leftrightarrow g}-\omega_c$ is the detuning between the cavity field 
mode resonance frequency $\omega_c$ and the TLS transition frequency $\omega_{e \leftrightarrow g}$. Finally, $g$ is the dipole coupling strength, which involves the normalized mode 
function of the single-mode radiation field and the transition dipole moment between $\ket{g}$ and $\ket{e}$.

%%%%%%%%%%%%%%%%%%%%%%%%%%%%%%%%%%%%
%%%%%%%%%%%%%%%%%%%%%%%%%%%%%%%%%%%%%%%

We suppose that at $t=0$ there are no correlations between the field and the TLS. Furthermore, we set the TLS to be initially in the excited state. Thus, our general initial quantum state reads
\begin{equation}\label{initialstate}
\ket{\psi(t=0)}=\ket{e}\otimes\sum_{n=0}^\infty a_n\ket{n},
\end{equation}
where $\ket{n}$ ($n\in {\mathbb N}_0$) are the normalized photon number states and $\sum_{n=0}^\infty\lvert{a_n}\rvert^2=1$. The time evolution is governed by the Schr\"odinger equation acting on
the initial state~(\ref{initialstate}) yields
\begin{equation}
\ket{\psi(t)}=\sum_{n=1}^{\infty}c_{e,n-1}(t)\ket{e,n-1}+c_{g,n}(t)\ket{g,n}, \label{3}
\end{equation}
where~\cite{Wolfgang}
\begin{eqnarray}
 c_{e,n-1}(t)&=&e^{-i\frac{\Delta t}{2}} \left[\cos(\lambda_n t)+i\frac{\Delta}{2 \lambda_n} \sin(\lambda_n t) \right] a_{n-1}, \nonumber \\
 c_{g,n}(t)&=&-i e^{i\frac{\Delta t}{2}} \frac{g \sqrt{n}}{\lambda_n} \sin(\lambda_n t) a_{n-1}, \nonumber
\end{eqnarray}
and where $\lambda_n=\sqrt{\Delta^2/4 + g^2 n}$ is the effective Rabi frequency. The state of the TLS upon emerging from the cavity is obtained by tracing out the state of the field,
\begin{eqnarray}
 \hat{\rho}(g,t) &=& \mathrm{Tr}_\mathrm{F}\{\ket{\psi(t)}\bra{\psi(t)}\}\nonumber\\
&=&\begin{bmatrix}
     a_{ee}(t) & a_{eg}(t) \\
     a^*_{eg}(t) & 1-a_{ee}(t) 
  \end{bmatrix}, \label{rhoatom1}
\end{eqnarray}
where
\begin{eqnarray}
 a_{ee}(t)&=& \sum_{n=1}^{\infty}|a_{n-1}|^2\bigg{[}\cos^2(\lambda_nt)+\frac{\Delta^2}{4\lambda_{n}^2}\sin^2(\lambda_{n}t)\bigg{]},  \\
 a_{eg}(t)&=&\sum_{n=1}^{\infty}a_na^*_{n-1}\bigg{[}\cos(\lambda_{n+1}t)+i\frac{\Delta}{2\lambda_{n+1}}\sin(\lambda_{n+1}t)\bigg{]} \nonumber \\
 &&\times\frac{ig\sqrt{n}}{\lambda_n} \sin(\lambda_nt) e^{-i\Delta t}.
\end{eqnarray}
In the next stage of the experiment, the TLS flies from the cavity to the detector. During this time spontaneous emission may occur. We include this effect in our calculations by using a simple 
Markovian description
\begin{equation}
 \frac{d \hat{\rho}}{dt}=\tfrac{\gamma}{2} \left(2 \hat{\sigma}_- \hat{\rho} \hat{\sigma}_+ - \hat{\sigma}_+ \hat{\sigma}_- \hat{\rho}- \hat{\rho}\hat{\sigma}_+ \hat{\sigma}_- \right), \label{sd}
\end{equation}
where $\gamma$ is the spontaneous emission rate of the TLS. Equation~\eqref{sd} is written in the frame rotating at the resonance frequency of the TLS. Two characteristic times enter our 
discussion: $\tau_c$, the duration of the matter--field interaction in the cavity, and $\tau_f$, the flying time from the cavity to the detector. The solution in Eq.~\eqref{rhoatom1} at $t=\tau_c$ 
can be considered
as the initial condition for Eq.~\eqref{sd}. Thus, the state of the TLS reaching the detector is
\begin{equation}
 \hat{\rho}(g)= \begin{bmatrix}
     a_{ee}(g,\tau_c) e^{-\gamma \tau_f} & a_{eg}(g,\tau_c)e^{-\gamma \tau_f/2} \\
     a^*_{eg}(g,\tau_c)e^{-\gamma \tau_f/2} & 1-a_{ee}(g,\tau_c)e^{-\gamma \tau_f} 
  \end{bmatrix}. \label{rhoatom2}
\end{equation}
Equation~\eqref{rhoatom2} yields a complete description of our setup and it applies to all the possible initial conditions of the field. A major theme of our subsequent discussion 
will be the analysis of Eq.~\eqref{rhoatom2} in the context of quantum estimation theory, where we shall seek optimal estimators for the coupling strength $g$.

\section{Quantum estimation theory}
\label{II}

In this section we summarize basic facts about parameter estimation in quantum theory which are relevant for our subsequent discussion and which have been reviewed in detail by Helstrom~\cite{Helstrom1976}. 
In particular, we summarize the methods and the concept behind them in order to provide an optimal estimation for the dipole coupling strength $g$ from the density matrix~\eqref{rhoatom2}.

The observational strategy for estimating $g$, a real number, can be expressed as a search for a POVM defined on the set $\Theta\subseteq\mathbb{R}$ of all possible values of $g$. 
The elements of the POVM represent the measurements to be performed on the TLS, which result in estimates $\tilde g$ of $g$, where $\tilde g$ is a random variable. The probability that 
it lies in a particular region $\Delta$ of the set $\Theta$, provided that the true value of the estimated dipole coupling is $g$, reads
\begin{equation}
 P\left(\tilde g \in \Delta| g \right)=\mathrm{Tr}\{\hat{\rho}(g) \hat{\Pi}(\Delta)\}.
\end{equation}
$\hat{\Pi}(\Delta)$ is an element of the POVM which is a mapping of regions $\Delta \subset \Theta$ into positive semidefinite operators on the Hilbert space $\mathbb{C}^2$ of the TLS with 
the following properties:
\begin{equation}
0 \leqslant \hat{\Pi}(\Delta) \leqslant \hat{I},\ \hat{\Pi}(\emptyset)=\hat{0},\ \text{and}\ \hat{\Pi}(\Theta)=\hat{I}, \label{POVM}
\end{equation}
where $\emptyset$ stands for the empty set, and $\hat{0}$ and $\hat{I}$ are the null and identity operators. Furthermore, we suppose that POVM elements on compact intervals $\Delta$ can be 
written as integrals with the help of the infinitesimal operators $d\hat{\Pi}(g)$, thus yielding
\begin{eqnarray}
 \hat{\Pi}(\Delta)=\int_{\Delta}d\hat{\Pi}(g)\quad\text{and}\quad\int_{\Theta}d\hat{\Pi}(g)=\hat{I}. \label{infinitesimal}
\end{eqnarray}
The conditional p.d.f. of the estimate $\tilde g$ is given by
\begin{equation}
 p(\tilde g | g)d \tilde g=\mathrm{Tr} \{\hat{\rho}(g) d\hat{\Pi}(\tilde g) \}, \label{condpdf}
\end{equation}
where $d \tilde g$ represents an infinitesimal compact interval in the set $\Theta$.

The \textit{Bayesian formulation} of the estimation problem seeks for the best estimator which minimizes the average cost of its application. In order to solve this estimation problem we have 
to provide an a priori p.d.f.\ $z(g)$ of $g$ to be estimated and a cost function $C(\tilde g, g)$, which asses the cost of error in the estimate. Now, combining the Bayesian estimation procedure 
with the strategy represented by the POVM in Eq.~\eqref{POVM} and including integral representation in Eq.~\eqref{infinitesimal}, we obtain for the average cost
\begin{equation}\label{averagecost}
\bar{C}=\mathrm{Tr} \left \{\int_{\Theta} dg \int_{\Theta} d\hat{\Pi}(\tilde g) z(g) C(\tilde g, g) \hat{\rho}(g) \right\}.
\end{equation}
We are looking for the $d\hat{\Pi}(\tilde{g})$ which minimizes $\bar C$. Our problem has thus been rephrased as a variational problem formulated on the space of all POVMs. In order to solve 
this problem one ought to first define $C(\tilde g, g)$. In this article we will employ the frequently used quadratic cost function
\begin{equation}
 C(\tilde g, g)=(\tilde g-g)^2, \label{qcostm}
\end{equation}
which leads to the minimum mean--square error (MMSE) estimator, and the delta-valued cost function
\begin{equation}
 C(\tilde g, g)=-\delta(\tilde g-g), \label{qcostd}
\end{equation}
which leads to the maximum likelihood (ML) estimator~\cite{Helstrom1976}.

In the following sections we will investigate both the MMSE and the ML estimation scenarios for different prior p.d.f.

\section{Minimum mean--square error estimator}
\label{III}

In the case of the quadratic cost function there is a more direct way to formulate the variational problem of Eq.~\eqref{averagecost}~\cite{Personick}. Let us assume that the elements of the 
POVM are projectors, with the infinitesimal operator
\begin{equation}
 d\hat{\Pi}(\tilde g)= \ket{\tilde g} \bra{\tilde g} d \tilde g, \nonumber 
\end{equation}
where $\ket{\tilde g}$ are the eigenstates of the estimator
\begin{equation}
 \hat{M}=\int_{\Theta} \tilde{g} d \hat{\Pi}(\tilde{g}) = \int_{\Theta} \tilde{g} \ket{\tilde{g}} \bra{\tilde{g}} d \tilde{g}, \label{MSE}
\end{equation}
and here $\tilde g$ stands for the all possible values of the estimate. In fact, the most convenient way to think about $\hat{M}$ is to consider it as an operator to be measured, with its 
eigenvalues the estimates of $g$~\cite{Zsolt}. Furthermore, the Hilbert space is $\mathbb{C}^2$ in our setup, which means that the POVM has only two projectors as elements, which project onto 
the eigenstates of $\hat{M}$. The average cost in Eq.~\eqref{averagecost}, together with Eq.~\eqref{MSE}, yields
\begin{equation}
 \bar C [\hat{M}]=\mathrm{Tr}\Big\{ \int_{\Theta} z(g) \bigl(\hat{M}-g \hat{I}\bigr)^2 \hat{\rho}(g) d g \Big\}. \label{Ctosolve}
\end{equation}
The unique Hermitian operator $\hat{M}_{\text{min}}$, the MMSE estimator, which minimizes $\bar C [\hat{M}]$ is the solution of the operator equation~\cite{Personick}
\begin{equation}
\hat{\Gamma}_0 \hat{M}_{\text{min}} +\hat{M}_{\text{min}} \hat{\Gamma}_0=2\hat{\Gamma}_1,
\label{eq:tosolve}
\end{equation}
where we have introduced the family of operators ($k=0,1,2$):
\begin{equation}
\hat{\Gamma}_k= \int_{\Theta} g^k z(g) \hat{\rho}(g) d g.
\label{eq:Gamma}
\end{equation}
The solution to Eq.~\eqref{eq:tosolve} reads
\begin{equation}
\hat{M}_{\text{min}}=2 \int^\infty_0 \exp(-\hat{\Gamma}_0 x) \hat{\Gamma}_1 \exp(-\hat{\Gamma}_0 x) d x,
\label{eq:intx}
\end{equation}
and the associated average minimum cost of error for the MMSE estimator is
\begin{equation}
 \bar C_{\text{min}}=\mathrm{Tr}\{\hat{\Gamma}_2-\hat{M}_{\text{min}} \hat{\Gamma}_0 \hat{M}_{\text{min}} \}. \label{mincost}
\end{equation}

In order to gain insight into the structure of $\hat{M}_{\text{min}}$, let us concentrate on resonant interactions $\Delta=\omega_{e \leftrightarrow g}-\omega_c=0$. We also consider the 
initial state of the single-mode field in \eqref{initialstate} to be the ground state, $a_0=1$. In this case Eq.~\eqref{rhoatom2} reads
\begin{equation}
 \hat{\rho}(g)=\begin{bmatrix}
     \cos^2(g \tau_c) e^{-\gamma \tau_f} & 0 \\
     0 & 1- \cos^2(g \tau_c)e^{-\gamma \tau_f} 
  \end{bmatrix}. \label{rhoatom3}
\end{equation}

We assume that the random variable $g$ to be estimated is characterized by its mean value $g_0$ and variance $\sigma^2$. In order to connect these parameters to experimental setups, we start with the 
position-dependent dipole coupling of the matter--field interaction~\cite{Wolfgang},
\begin{equation}
 g(\vec{r}_q)=-\sqrt{\frac{\hbar \omega_c}{2 \epsilon_0}}\bra{g} \hat{\vec{d}} \ket{e}\cdot \vec{u}(\vec{r}_q)/\hbar, \nonumber
\end{equation}
where $\hat{\vec{d}}$ is the dipole operator, $\epsilon_0$ the permittivity of vacuum, and $\vec{r}_q$ the position vector. The normalized mode function of the single-mode 
radiation field, $\vec{u}(\vec{r})$, is a solution to the Helmholtz equation and fulfills the Coulomb gauge and the cavity boundary conditions. However, every passing TLS also experiences 
changes in the dipole coupling due to the waist of the field mode. Experimental studies usually integrate the collected data over the flying time through the cavity and thus obtain an average 
coupling strength $g_0$; cf., for example, Ref.~\cite{Gleyzes}. This method results also in a variance $\sigma^2$ of the measured coupling strength. In the following, we are going to discuss two 
prior p.d.f.\ whose mean values and variances coincide with the values defined here.

\subsection{Gaussian probability density function}

In this subsection we consider $\Theta=\mathbb{R}$ and the prior p.d.f.
\begin{equation}\label{gaussianprior}
z(g)=\frac{1}{\sqrt{2\pi\sigma^2}}e^{-\frac{(g-g_0)^2}{2\sigma^2}}, \quad g \in \Theta. 
\end{equation}
As $z(g)$ and the density matrix in Eq.~\eqref{rhoatom3} are given, the operators defined in Eq.~\eqref{eq:Gamma} can be evaluated explicitly, yielding
\begin{eqnarray}
 \hat{\Gamma}_0&=&\begin{bmatrix}
     a e^{-\gamma \tau_f} & 0 \\
     0 & 1- a e^{-\gamma \tau_f} 
  \end{bmatrix}, \nonumber\\
 a&=&\frac{1+e^{-2 \sigma^2 \tau^2_c}\cos(2 g_0 \tau_c)}{2}, \nonumber \\
 \hat{\Gamma}_1&=&\begin{bmatrix}
     b e^{-\gamma \tau_f} & 0 \\
     0 & g_0- b e^{-\gamma \tau_f} 
  \end{bmatrix}, \label{abgauss} \nonumber\\
  b&=&\frac{g_0+e^{-2 \sigma^2 \tau^2_c} \left[ g_0\cos(2 g_0 \tau_c)-2 \sigma^2 \tau_c \sin(2 g_0 \tau_c) \right]}{2}, \nonumber \\
 \hat{\Gamma}_2&=&\begin{bmatrix}
     c e^{-\gamma \tau_f} & 0 \\
     0 & g^2_0+\sigma^2- c e^{-\gamma \tau_f} 
  \end{bmatrix}, \nonumber
\end{eqnarray}
and
\begin{multline}
  c=\frac{(g^2_0+\sigma^2)\left[1+e^{-2 \sigma^2 \tau^2_c}\cos(2 g_0 \tau_c)\right]}{2} \\
  -2g_0 \sigma^2 \tau_c e^{-2 \sigma^2 \tau^2_c} \sin(2 g_0 \tau_c)\\
  -2 \sigma^4 \tau^2_c e^{-2 \sigma^2 \tau^2_c} \cos(2 g_0 \tau_c). \nonumber 
\end{multline}
Now, Eq.~\eqref{eq:intx} can be directly calculated and the MMSE estimator reads
\begin{equation}
\hat{M}_{\text{min}}=\begin{bmatrix}
     \frac{b}{a} & 0 \\
     0 & \frac{g_0- b e^{-\gamma \tau_f}}{1- a e^{-\gamma \tau_f}} 
  \end{bmatrix}.  \label{eq:Moperator} 
\end{equation}
The average minimum cost of error is
\begin{multline}
 \bar C_{\text{min}}=g^2_0+\sigma^2-\left(\frac{g_0- b e^{-\gamma \tau_f}}{1- a e^{-\gamma \tau_f}} \right)^2 \\
 -a e^{-\gamma \tau_f} \left[\frac{b^2}{a^2}-\left(\frac{g_0- b e^{-\gamma \tau_f}}{1- a e^{-\gamma \tau_f}} \right)^2\right]. \nonumber
\end{multline}

To illustrate the meaning of the MMSE estimator $\hat{M}_{\text{min}}$ and the average minimum cost of error $\bar C_{\text{min}}$ we consider a situation where the experimentalist, 
based on their prior expectations of the coupling strength $g$, sets the duration of the matter--field interaction $\tau_c=\pi/(2g_0)$. This reflects the fact that the experimentalist expects 
the TLS to emit a photon into the field mode and fly towards the detectors in its ground state. This setup yields
\begin{equation}
 \hat{M}_{\text{min}}=\begin{bmatrix}
     g_0 & 0 \\
     0 & g_0 
  \end{bmatrix}, \quad \bar C_{\text{min}}=\sigma^2, \nonumber
\end{equation}
which means that the estimates $\tilde g$ are always $g_0$ no regardless of the applied projective measurement. Furthermore, the average minimum cost of error is $\sigma^2$. 
Thus, this scenario simply reinforces prior expectations on the true value of $g$. Another inconclusive setup would be when $\tau_c=\pi/g_0$, i.e., the experimentalist expects that the 
TLS will not emit a photon into the field mode.

A much more interesting scenario is when $\tau_c=\pi/(4g_0)$ or in other words the experimentalist expects the TLS to emit a photon with $50\%$ probability. Now, we have
\begin{equation}
 \hat{M}_{\text{min}}=\begin{bmatrix}
     g_0-\frac{\sigma^2 \pi}{2 g_0}e^{-\frac{\pi^2}{8}\frac{\sigma^2}{g^2_0}} & 0 \\
     0 & g_0+\frac{\sigma^2 \pi}{2 g_0} \frac{1}{2 e^{\gamma \tau_f}-1}e^{-\frac{\pi^2}{8}\frac{\sigma^2}{g^2_0}} \nonumber
  \end{bmatrix},
\end{equation}
and
\begin{equation}
 \bar C_{\text{min}}=\sigma^2-\frac{\sigma^4 \pi^2}{4 g^2_0}\frac{1}{2 e^{\gamma \tau_f}-1}e^{-\frac{\pi^2}{4}\frac{\sigma^2}{g^2_0}}. \nonumber
\end{equation}
Measuring the TLS in the excited state results the estimate
\begin{equation}
 \tilde g=g_0-\frac{\sigma^2 \pi}{2 g_0}e^{-\frac{\pi^2}{8}\frac{\sigma^2}{g^2_0}}, \nonumber
\end{equation}
with probability
\begin{equation}
 p=\cos^2\left(\frac{\pi}{4} \frac{g}{g_0}\right) e^{-\gamma \tau_f}. \nonumber
\end{equation}
The destructive effects of the spontaneous decay are revealed here, because when $\gamma \tau_f \gg 1$ this probability reduces to zero and therefore the measurement cannot obtain the estimate belonging
to the excited state of the TLS. When the measurement yields the other outcome, the state is projected onto the ground state of the TLS, and the resulting estimate is
\begin{equation}
 \tilde g=g_0+\frac{\sigma^2 \pi}{2 g_0} \frac{1}{2 e^{\gamma \tau_f}-1}e^{-\frac{\pi^2}{8}\frac{\sigma^2}{g^2_0}} \nonumber
\end{equation}
with probability
\begin{equation}
 p=1-\cos^2\left(\frac{\pi}{4} \frac{g}{g_0}\right) e^{-\gamma \tau_f}. \nonumber
\end{equation}
When $\gamma \tau_f \gg 1$, this result is obtained with certainty, and the resulting estimate is simply $g_0$ and $\bar C_{\text{min}}=\sigma^2$. Again our prior expectations of the true value of $g$ 
are reinforced. In general, the average estimate is
\begin{eqnarray}
\mathrm{E}[\tilde g\,|g]&=&\mathrm{Tr}\{\hat{M}_\text{min}\hat{\rho}(g)\} \nonumber \\
&=&g_0-\cos\left(\frac{\pi}{2} \frac{g}{g_0}\right)\frac{\sigma^2 \pi}{2 g_0} \frac{1}{2 e^{\gamma \tau_f}-1}e^{-\frac{\pi^2}{8}\frac{\sigma^2}{g^2_0}}, \nonumber
\end{eqnarray}
which is conditioned on the true value of $g$. Performing several measurements with identical TLSs yields an average estimate from which one may deduce the value of $g$. 
When the standard deviation $\sigma$ of the prior p.d.f.\ is set very large compared to the prior mean $g_0\ll\sigma$, we allow the true value of $g$ to be far from the prior mean. 
In this context the estimates turn out to be again $g_0$ and accordingly the average minimum cost of error is $\sigma^2$. In the case when the true value of $g$ is $g_0$, we find 
$\mathrm{E}[\tilde g\,|g_0]=g_0$.

In the next step, the accuracy with which $g$ can be estimated is characterized by the mean--squared error $\mathrm{E}\bigl[(\tilde g -g)^2|g\bigr]$~\cite{commentonus}. 
The lower bound of the mean--squared error 
is given by a quantum Cram\'er--Rao-type inequality~\cite{Helstrom1976}
\begin{equation}
\mathrm{E}\bigl[(\tilde g -g)^2|g\bigr]=\mathrm{Tr}\bigl\{\bigl(\hat{M}_{\text{min}}-g \hat{I}\bigr)^2 \hat{\rho}(g) \bigr\} \geqslant
\frac{|x'(g)|^2}{\mathrm{Tr}\big\{\hat{\rho}(g)\hat{L}^2 \big\}},
\label{eq:lbound}
\end{equation}
where
\begin{equation}
 x'(g)= \mathrm{Tr}\{\hat{M}_{\text{min}} \frac{\partial}{\partial g} \hat{\rho}(g)\}, \nonumber
\end{equation}
and the symmetrized logarithmic derivative $\hat{L}$ of the density matrix $\hat{\rho}(g)$ is defined as
\begin{equation}
\frac{\partial \hat{\rho}(g)}{\partial g}=\frac{1}{2}\left[\hat{L}\hat{\rho}(g)+\hat{\rho}(g) \hat{L}\right]. \nonumber
\end{equation}
If we consider the spectral decomposition 
\begin{equation}
\hat{\rho}(g)= \cos^2(g \tau_c) e^{-\gamma \tau_f} \ket{e} \bra{e}+\left(1- \cos^2(g \tau_c)e^{-\gamma \tau_f}\right)\ket{g} \bra{g}, \nonumber
\end{equation}
then
\begin{equation}
\hat{L}=-2 \tau_c \tan(g \tau_c) \ket{e} \bra{e}+\tau_c \frac{\sin(2 g \tau_c) e^{-\gamma \tau_f}}{1- \cos^2(g \tau_c)e^{-\gamma \tau_f}}\ket{g} \bra{g}. \nonumber
\end{equation}
Hence, we have
\begin{eqnarray}
\mathrm{E}\left[(\tilde g -g)^2|g\right] &\geqslant& \left[1- \cos^2(g \tau_c)e^{-\gamma \tau_f}\right] \label{explbound} \\
&\times& \cos^2(g \tau_c)e^{-\gamma \tau_f} \frac{\left(g_0-b/a\right)^2}{\left(1-a e^{-\gamma \tau_f}\right)^2}. \nonumber
\end{eqnarray}
In the inconclusive cases when the experimentalist sets the interaction times either to $\pi/(2g_0)$ or $\pi/g_0$ the inequality in Eq.~\eqref{explbound} yields
\begin{equation}
\mathrm{E}\left[(\tilde g -g)^2|g\right]\geqslant 0, \nonumber
\end{equation}
which also means that when we bolster our prior knowledge then the lower bound of the accuracy is the smallest. Now, for the interesting case of $\tau_c=\pi/(4g_0)$, we find
\begin{eqnarray}
&&\mathrm{E}\left[(\tilde g -g)^2|g\right] \geqslant \left[1- \cos^2\left(\frac{\pi}{4}\frac{g}{g_0}\right)e^{-\gamma \tau_f}\right] \nonumber \\
&&\times \cos^2\left(\frac{\pi}{4}\frac{g}{g_0}\right)e^{-\gamma \tau_f} \frac{\frac{\pi^4}{16}\frac{\sigma^4}{g^4_0}e^{-\frac{\pi^2}{4}\frac{\sigma^2}{g^2_0}}}{\left(2-e^{-\gamma \tau_f}\right)^2}. 
\end{eqnarray}
It is worth noting that the in inconclusive situation $\gamma \tau_f \gg 1$, when the estimate of the 
coupling strength is $g_0$, the lower bound of the 
mean--squared error decreases also towards zero. This fact is in agreement with the inconclusive scenarios where $\tau_c=\pi/(2g_0)$ and $\tau_c=\pi/g_0$, where the left-hand side of the quantum 
Cram\'er--Rao inequality is zero, the minimum allowed value. It seems in the context of our system that the extremal behaviors of lower bounds on the accuracy are related only to inconclusive 
estimation scenarios.

\subsection{Uniform probability density function}

In this subsection we consider a uniform prior p.d.f. As the only prior knowledge about the coupling $g$ is its mean value $g_0$ and variance $\sigma^2$, we set the parameter space 
$\Theta=[g_0-\sqrt{3} \sigma,g_0+\sqrt{3} \sigma ]$ and p.d.f.
\begin{equation}
 z(g)=\frac{1}{2\sqrt{3}\sigma}, \quad g \in \Theta. \label{uniformaprior}
\end{equation}
Similarly to the previous subsection we determine the operators defined in Eq.~\eqref{eq:Gamma}
\begin{eqnarray}
 &&\hat{\Gamma}_0=\begin{bmatrix}
     a' e^{-\gamma \tau_f} & 0 \\
     0 & 1- a' e^{-\gamma \tau_f} 
  \end{bmatrix}, \nonumber \\ 
  &&a'=\frac{1}{2}+\frac{\sin(2\sqrt{3}\sigma \tau_c)\cos(2g_0 \tau_c)}{4\sqrt{3}\sigma \tau_c}; \nonumber \\
  &&\hat{\Gamma}_1=\begin{bmatrix}
     b' e^{-\gamma \tau_f} & 0 \\
     0 & g_0- b' e^{-\gamma \tau_f} 
  \end{bmatrix}, \label{abuniform} \\
  &&b'=\frac{g_0}{2} - \frac{ \sin(2g_0\tau_c)\sin(2 \sqrt{3} 
\sigma \tau_c)}{8\sqrt{3} \sigma \tau_c^2} \nonumber \\
   &&+\frac{ \sqrt{3} \sigma \sin(2 g_0 
\tau_c)\cos(2\sqrt{3}\sigma\tau_c)+g_0\cos(2 g_0\tau_c)\sin(2 \sqrt{3} \sigma \tau_c)}{4 \sqrt{3} \sigma \tau_c}, \nonumber 
\end{eqnarray}
and
\begin{eqnarray}
 &&\hat{\Gamma}_2=\begin{bmatrix}
     c' e^{-\gamma \tau_f} & 0 \\
     0 & g^2_0+\sigma^2- c' e^{-\gamma \tau_f} 
  \end{bmatrix}, \nonumber \\
  &&c'=\frac{g^2_0+\sigma^2}{2}+\frac{(g^2_0+3 \sigma^2)\sin(2\sqrt{3}\sigma \tau_c)\cos(2g_0 \tau_c)}{4 \sqrt{3} \sigma \tau_c} \nonumber \\
  &&+\frac{\sqrt{3}\sigma \cos(2\sqrt{3}\sigma \tau_c)\cos(2g_0 \tau_c)-g_0\sin(2\sqrt{3}\sigma \tau_c)\sin(2g_0 \tau_c)}{4 \sqrt{3} \sigma \tau^2_c} \nonumber \\
  &&-\frac{\sin(2\sqrt{3}\sigma \tau_c)\cos(2g_0 \tau_c)}{8 \sqrt{3} \sigma \tau^3_c}+\frac{g_0\sin(2g_0\tau_c)\cos(2\sqrt{3}\sigma \tau_c)}{2 \tau_c}. \nonumber
\end{eqnarray}
As the structure of the operators $\hat{\Gamma}_k$ ($k=0,1,2$) is the same as in the previous subsection, where we have considered the Gaussian p.d.f., we obtain for the MMSE estimator
\begin{equation}
\hat{M}_{\text{min}}=\begin{bmatrix}
     \frac{b'}{a'} & 0 \\
     0 & \frac{g_0- b' e^{-\gamma \tau_f}}{1- a' e^{-\gamma \tau_f}} 
  \end{bmatrix}. \nonumber 
\end{equation}
The average minimum cost of error is
\begin{eqnarray}
 \bar C_{\text{min}}&=&g^2_0+\sigma^2-\left(\frac{g_0- b' e^{-\gamma \tau_f}}{1- a' e^{-\gamma \tau_f}} \right)^2 \nonumber \\
 &-& a' e^{-\gamma \tau_f} \left[\frac{b'^2}{a'^2}-\left(\frac{g_0- b' e^{-\gamma \tau_f}}{1- a' e^{-\gamma \tau_f}} \right)^2\right]. \nonumber
\end{eqnarray}
The two cases discussed $\tau_c=\pi/(2g_0)$ and $\tau_c=\pi/g_0$ were found to be inconclusive in the previous subsection. It is immediate to see from the structure of $\hat{\Gamma}_k$ 
that for a uniform prior p.d.f.\ these cases are not indecisive any more. Thus, supposing that nothing is known in advance about the true value of $g$ in the interval 
$[g_0-\sqrt{3} \sigma,g_0+\sqrt{3} \sigma ]$ actually reduces the number of inconclusive scenarios. Let us also reconsider $\tau_c=\pi/(4g_0)$, i.e., the experimentalist expects the TLS to emit a 
photon with $50\%$ probability, which was seen to be an interesting case of the previous subsection. The MMSE estimator is, in this case,
\begin{eqnarray}
 \hat{M}_{\text{min}}=\begin{bmatrix}
     g_0\left(1+x\right) & 0 \\
     0 & g_0\left(1- \frac{x}{2 e^{\gamma \tau_f}-1}\right)
  \end{bmatrix}, \nonumber
\end{eqnarray}
with
\begin{equation}
 x=\frac{2}{\pi} \cos \left(\frac{\sqrt{3}\pi}{2}\frac{\sigma}{g_0}\right)-\frac{4}{\sqrt{3}\pi^2}\frac{g_0}{\sigma} \sin \left(\frac{\sqrt{3}\pi}{2}\frac{\sigma}{g_0}\right). \nonumber
\end{equation}
The average minimum cost of error is
\begin{eqnarray}
 \bar C_{\text{min}}=\sigma^2-g^2_0 \frac{x^2}{2 e^{\gamma \tau_f}-1}. \nonumber
\end{eqnarray}
Measuring the TLS in the excited state results in the estimate
\begin{equation}
 \tilde g=g_0\left(1+x\right), \nonumber
\end{equation}
with probability
\begin{equation}
 p=\cos^2\left(\frac{\pi}{4} \frac{g}{g_0}\right) e^{-\gamma \tau_f}. \nonumber
\end{equation}
Once again we find that when $\gamma \tau_f \gg 1$ this probability reduces to zero and therefore a measurement cannot yield this estimate. Finding the TLS in the ground state results in the estimate
\begin{equation}
 \tilde g=g_0\left(1- \frac{x}{2 e^{\gamma \tau_f}-1}\right), \nonumber
\end{equation}
with probability
\begin{equation}
 p=1-\cos^2\left(\frac{\pi}{4} \frac{g}{g_0}\right) e^{-\gamma \tau_f}. \nonumber
\end{equation}
The situation is the same as that for the Gaussian prior p.d.f., i.e., when $\gamma \tau_f \gg 1$ one measures the TLS to be in the ground state with certainty, the estimate is simply $g_0$, and 
$\bar C_{\text{min}}=\sigma^2$. In any case, the average estimator is
\begin{equation}
\mathrm{E}[\tilde g\,|g]=g_0+g_0 x \frac{2\cos^2\left(\frac{\pi}{4} \frac{g}{g_0}\right)-1}{2 e^{\gamma \tau_f}-1}. \nonumber
\end{equation}
We note again the case when the true value of $g$ is $g_0$, then $\mathrm{E}[\tilde g\,|g_0]=g_0$. If $\gamma \tau_f \gg 1$ then the average estimator is also $g_0$ no matter what the true value of $g$ is; 
this is again an inconclusive scenario.

With the uniform prior p.d.f., the quantum Cram\'er--Rao inequality is
\begin{eqnarray}
&&\mathrm{E}\left[(\tilde g -g)^2|g\right] \geqslant  \left[1- \cos^2(g \tau_c)e^{-\gamma \tau_f} \right] \nonumber \\
&& \times \cos^2(g \tau_c)e^{-\gamma \tau_f} \frac{\left(g_0-b'/a'\right)^2}{\left(1-a' e^{-\gamma \tau_f}\right)^2}, \label{ulbound}
\end{eqnarray}
which yields, when $\tau_c=\pi/(4g_0)$,
\begin{eqnarray}
&&\mathrm{E}\left[(\tilde g -g)^2|g\right] \geqslant\left[1- \cos^2\left(\frac{\pi}{4}\frac{g}{g_0}\right)e^{-\gamma \tau_f} \right] \\
&& \cos^2\left(\frac{\pi}{4}\frac{g}{g_0}\right)e^{-\gamma \tau_f} \frac{4 g^2_0 
 x^2}{\left(2-e^{-\gamma \tau_f}\right)^2}. \nonumber 
\end{eqnarray}
The next subsection focuses on numerical simulations in order to understand the role of the detuning $\Delta$ and an initial field state with mean photon number larger than zero. We will investigate 
the deviations from the analytical results of this section and understand the changes inflicted on the estimates, the minimum average cost of error, and the left-hand side of the quantum Cram\'er--Rao inequality.

\subsection{Numerical results}

In the previous subsections we have calculated analytically the MMSE estimators for both the Gaussian~\eqref{gaussianprior} and the uniform~\eqref{uniformaprior} p.d.f. We have presented the simplest 
scenario, where the cavity field mode is initially in the ground state, $a_0=1$, which led to a diagonal form of the density matrix~\eqref{rhoatom3}. Furthermore, we have considered the single-mode 
field to be in resonance with the TLS transition, $\Delta=\omega_{e\leftrightarrow g}-\omega_c=0$, which has allowed us to perform the integrations in Eq.~\eqref{eq:Gamma}. Here, we show the numerical 
results obtained in more general cases, where the initial state of the field mode is a more general coherent state $\ket{\alpha}$, and where we may have non-zero detuning $\Delta \neq 0$. The coherent 
state is defined through the parameter $\alpha$~\cite{Perelomov86},
\begin{equation}
  \label{coherent}
 \ket{\alpha}=\sum_{n=0}^\infty 
  e^{-\frac{|\alpha|^2}{2}}
 \frac{\alpha^n}{\sqrt{n!}}
 \ket n,
  \quad\alpha=|\alpha|e^{i\phi},
\end{equation}
where $\ket{n}$ ($n\in {\mathbb N}_0$) are the photon number states and $\phi$ is the complex phase of $\alpha$; the mean photon number of this coherent state is $|\alpha|^2$. Here, we set $\phi=0$. 

\begin{figure}
\centering
\begin{subfigure}{\linewidth}% width of left subfigure
\includegraphics[width=8cm]{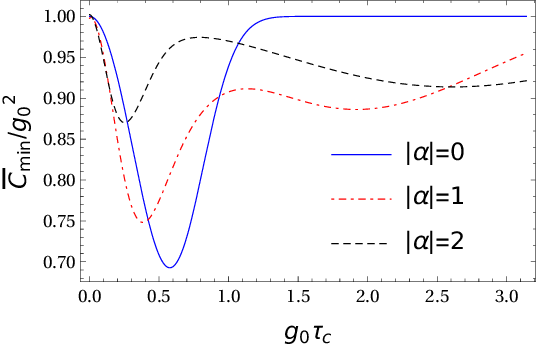}
\caption{The average minimum cost of error $\bar{C}_{\text{min}}/g_0^2$ as a function of $g_0\tau_c$}\label{fig:CostGauss}
\end{subfigure}
\begin{subfigure}{\linewidth}
\includegraphics[width=8cm]{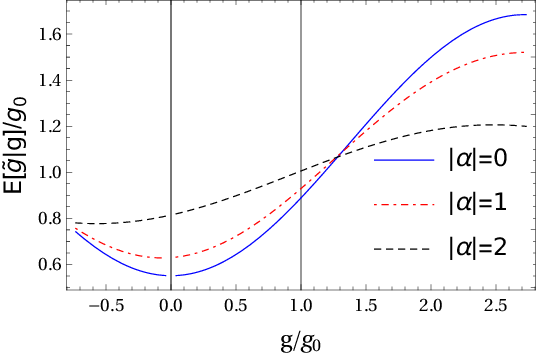}
\caption{The average estimator $E[\tilde{g}|g]/g_0$ as a function of $g/g_0$}\label{fig:avestimatorgauss}
\end{subfigure}
\begin{subfigure}{\linewidth}
\includegraphics[width=8cm]{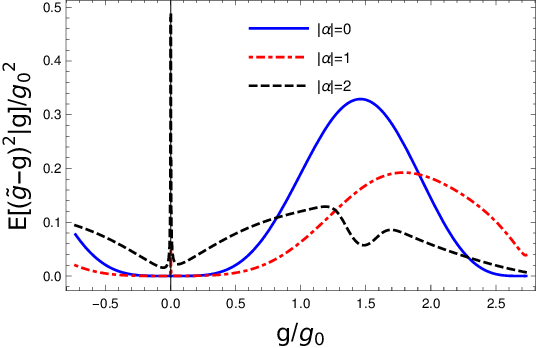}
\caption{The lower bound of the mean--squared error as a function of $g/g_0$}\label{fig:MSEGauss}
\end{subfigure}
\caption{Figures obtained for a Gaussian prior p.d.f., with mean $g_0$ and variance $\sigma^2/g^2_0=1$. We set $\gamma\tau_f=0$, such that no spontaneous decay occurs. 
In Fig. (a) each curve has a global minimum that decreases and shifts to larger values of $g_0\tau_c$ with increasing $|\alpha|$. At the time, when $\bar{C}$ attains its minimum, Fig. (b) 
displays biased average estimators, where the mean value $g_0$ of the prior p.d.f. is depicted by a vertical line. The curves in Fig. (c) characterizing the accuracy of the estimation scenario
have to be considered together with the appropriate curves in Fig. (b) in order to obtain a more complete information abut the optimal MMSE estimator.} \label{Figextra}
\end{figure}

{\it Gaussian p.d.f.\ and resonant interaction $\Delta=0$.}---The two parameters of the Gaussian p.d.f.\ are its mean $g_0$ and variance $\sigma^2$. To simplify the analysis we set $\gamma\tau_f=0$, 
so that no spontaneous emission may occur. We start our analysis with the simplest case $\alpha=0$. First of all we discuss the eigenvalues, i.e, the estimates, of the operator $\hat{M}_{\text{min}}$. 
One of the eigenvalues of $\hat{M}_{\text{min}}$ has a discontinuity at $\tau_c=0$. This can be shown by explicitly taking the limit
\begin{equation}
\lim_{\tau_c\to 0^+}\frac{g_0-b}{1-a}=\frac{g_0(3\sigma^2 + g_0^2)}{\sigma^2 + g_0^2},
\end{equation}
with $a$ and $b$ defined in Eq.~\eqref{abgauss}. For $\tau_c=0$ the function $\frac{g_0-b}{1-a}$ is not defined and the eigenvalue can be obtained only by starting again the whole calculation 
from Eq.~\eqref{rhoatom3}. The other eigenvalue is continuous and its value tends to $g_0$. At $\tau_c=0$ the eigenvalues of $\hat{M}_{\text{min}}$ are $g_0$ and $0$. This is simply due to the fact 
that no interaction occurred. Thus, estimates give either the prior expected coupling value or no coupling at all. When $\tau_c$ tends to infinity both curves approach the prior expected value
$g_0$, and the measurement is again inconclusive. In Fig.~\ref{fig:CostGauss} the average minimum cost of error $\bar{C}_{\text{min}}$ is plotted. 
At $\tau_c=0$ we find $\bar{C}_{\text{min}}=\sigma^2$, 
equal to the prior variance. The plot shows that there is global minimum of $\bar{C}_{\text{min}}$, which defines the recommended value of $g_0 \tau_c$ for the experimental detection. For a 
fixed value of $g_0$, we denote the recommended interaction time as $\tau_c^*$. 
%%%%%%%%%%%%%%%%%%%%%%%%%%%%%%%%%%%%%%%%%%%%%%%%%%%%%%%%%%%%%%%%%%%%%%%%%%%%%%%%%%%%%%%%%%%%%%%%%%%%%%%%%%%%%%
\begin{figure}
\includegraphics[width=8cm]{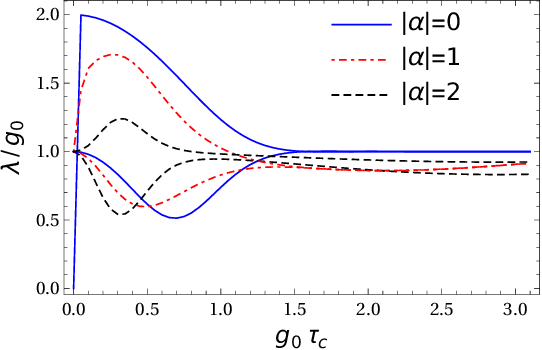}
\caption{The eigenvalues of $\hat{M}_{\text{min}}/g_0$ as a function of $g_0\tau_c$ in the case of the Gaussian prior p.d.f., with mean $g_0$ and variance $\sigma^2/g^2_0=1$. In the case $\alpha=0$ 
the two initial eigenvalues are $0$ and $g_0$. When $|\alpha|>0$, the eigenvalues are plotted from $\tau_c=0^+$, at which time they are equal to $g_0$. 
For large values of $\tau_c$, the eigenvalues tend to the same value $g_0$. 
We set $\gamma\tau_f=0$, such that no spontaneous decay occurs.}
\label{fig:EigenvaluesMtGauss}
\end{figure}
%%%%%%%%%%%%%%%%%%%%%%%%%%%%%%%%%%%%%%%%%%%%%%%%%%%%%%%%%%%%%%%%%%%%%%%%%%%%%%%%%%%%%%%%%%%%%%%%%%%%%%%%%%%%%

%%%%%%%%%%%%%%%%%%%%%%%%%%%%%%%%%%%%%%%%%%%%%%%%%%%%%%%%%%%%%%%%%%%%%%%%%%%%%%%%%%%%%%%%%%%%%%%%%%%%

%%%%%%%%%%%%%%%%%%%%%%%%%%%%%%%%%%%%%%%%%%%%%%%%%%%%%%%%%%%%%%%%%%%%%%%%%%%%%%%%%%%%%%%%%%%%%%%%%%%%

Finally, let us discuss the scenarios with finite field amplitude $|\alpha|$ i.e., the initial average photon number becomes non--zero. Now, we have to focus completely on numerical solutions, because
analytical calculations are not possible. The numerical results in Fig.~\ref{fig:EigenvaluesMtGauss} show the eigenvalues of $\hat{M}_{\text{min}}$. 
Contrary to the behaviour encountered for $\alpha=0$, here all eigenvalues seem to start from $g_0$. However, this is true only for $\tau_c=0^+$. When $\tau_c=0$ the eigenvalues are $0$ and $g_0$, but we 
can not obtain them due to the finite numerical sum of the field amplitude. Starting from $\tau_c=0^+$ the eigenvalues 
are robust against the increase of the terms in the summation provided that the numerical normalization of the coherent state is larger than $0.99$.
Similarly to the case $\alpha=0$ the eigenvalues 
are approaching $g_0$ as $\tau_c\to\infty$. The average minimum cost error $\bar{C}_{\text{min}}$ starts for all the values of $|\alpha|$ at $\bar{C}_{\text{min}}=\sigma^2$ and reaches a global minimum for $\tau_c=\tau_c^*.$
This value depends on $|\alpha|$ and decreases with increasing $|\alpha|$. As more photons are involved in the 
interaction, i.e., the TLS and the field mode undergo many exchanges of photons, more information gets lost in the different photon number states $\ket{n}$. Therefore, the lowest average minimum 
cost of error is obtained when the field is in the vacuum state, $\ket{\alpha}=\ket{0}$. However, more photons in the interaction result in the appearance of higher Rabi frequencies $g \sqrt{n}$,
which in turn means that the minimum value is reached quicker. We note that different $\alpha$ with the same absolute value show the same behavior both for the eigenvalues of $\hat{M}_{\text{min}}$ 
and $\bar{C}_{\text{min}}$, thus the mean photon number is the only significant variable for the estimation of the dipole coupling strength.

Next, we calculate the average estimator $E[\tilde{g}|g]$, which is determined from the measurement data and from which the value of $g$ can be deduced. Repeated measurements of $\hat{M}_{\text{min}}$ 
at $\tau_c=\tau_c^*$ give different outcomes whose average is related with the true value $g$. Fig.~\ref{fig:avestimatorgauss} shows some curves for different values of $\alpha$ which clearly 
demonstrate that the obtained MMSE estimator is biased. Furthermore, using Eq.~\eqref{eq:lbound} we plot in Fig.~\ref{fig:MSEGauss} the lower bounds of 
the mean--squared error.
In the case $g=g_0$, the lower bound of the mean--squared error decreases whenever $|\alpha|\neq 0$. By taking into account 
the behavior of the average estimate $E[\tilde{g}|g]$, which at $g=g_0$ approaches the value of $g_0$ with increasing $|\alpha|$ (see Fig.~\ref{fig:avestimatorgauss}), we can conclude 
that increasing values of $|\alpha|$ lead to measurement strategies which reinforce our prior expectations.

\begin{figure}
\includegraphics[width=8cm]{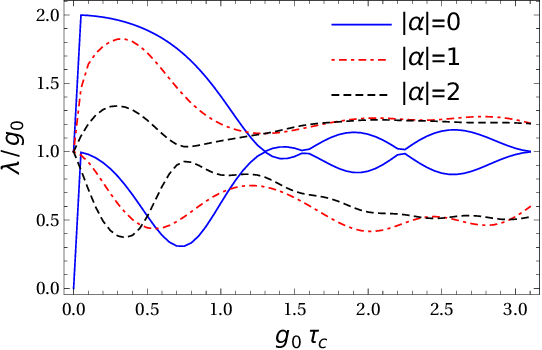}
\caption{The eigenvalues of $\hat{M}_{\text{min}}/g_0$ as a function of $g_0\tau_c$ in the case of the uniform prior p.d.f.\ with mean $g_0$ and variance $\sigma^2/g^2_0=1$. 
In the case $\alpha=0$ the two initial eigenvalues are $0$ and $g_0$. The eigenvalues are plotted from $\tau_c=0^+$ for all $|\alpha|>0$ and their starting values are $g_0$. 
For large values of $g_0\tau_c$, the eigenvalues tend 
to the same value $g_0$, but slower than in Fig.~\ref{fig:EigenvaluesMtGauss}. We set $\gamma\tau_f=0$.}
\label{fig:EigenvaluesMtUniform}
\end{figure}

\begin{figure}
\centering
\begin{subfigure}{\linewidth}% width of left subfigure
\includegraphics[width=8cm]{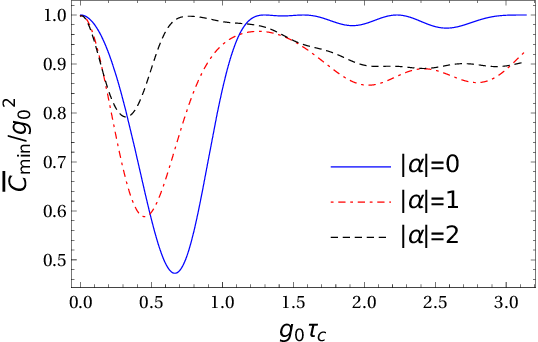}
\caption{The average minimum cost of error $\bar{C}_{\text{min}}/g_0^2$ as a function of $g_0\tau_c$.}\label{costuniform}
\end{subfigure}
\begin{subfigure}{\linewidth}
\includegraphics[width=8cm]{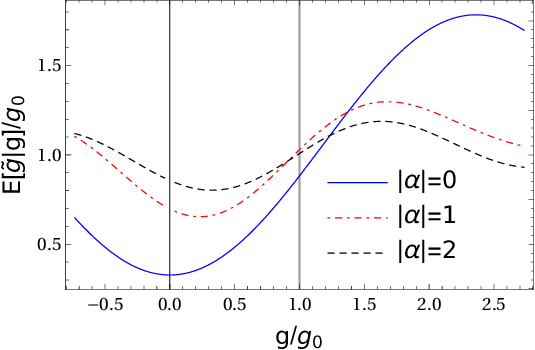}
\caption{The average estimator $E[\tilde{g}|g]/g_0$ as a function of $g/g_0$.}\label{fig:avestimatoruniform}
\end{subfigure}
\begin{subfigure}{\linewidth}
\includegraphics[width=8cm]{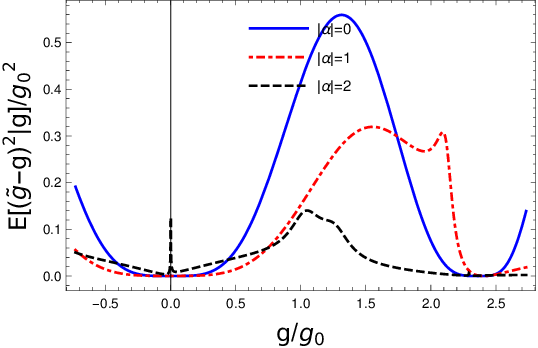}
\caption{The lower bound of the mean--squared error as a function of $g/g_0$.}\label{fig:MSEUniform}
\end{subfigure}
\caption{Figures obtained for a uniform prior p.d.f. The parameters are set to the same value as in Fig. \ref{Figextra}. The curves display very similar properties to those 
corresponding in Fig. \ref{Figextra}.}
\end{figure}

{\it Uniform prior p.d.f.\ and resonant interaction $\Delta=0$.}---The two parameters of the uniform p.d.f.\ are again the mean $g_0$ and the variance $\sigma^2$. We assume again that no spontaneous
emission occurs, i.e., $\gamma\tau_f=0$. In Fig.~\ref{fig:EigenvaluesMtUniform} the measurement estimates, or the eigenvalues of the MMSE operator $\hat{M}_{\text{min}}$, are shown. 
If $\alpha=0$, the eigenvalues show a discontinuity around at $\tau_c=0$, as in the case of Gaussian p.d.f., which can be seen from the analytical calculation of $\hat{M}_{\text{min}}$. 
The corresponding limit reads
\begin{equation}
\lim_{\tau_c\to 0^+}\frac{g_0-b'}{1-a'}=\frac{g_0(3\sigma^2 + g_0^2)}{\sigma^2 + g_0^2},
\end{equation}
with $a'$ and $b'$ defined in \eqref{abuniform}.
The average cost function $\bar{C}_{\text{min}}$ plotted in Fig.~\ref{costuniform} starts from the prior variance $\sigma^2$ and after reaching a global minimum approaches again the prior variance as $\tau_c\to\infty$. 
Fig.~\ref{fig:avestimatoruniform} shows the average estimator $E[\tilde{g}|g]$ at the time $\tau_c^*$ when $\bar{C}_{\text{min}}$ attains its minimum. The lower bound of the 
mean--squared error is shown in Fig.~\ref{fig:MSEUniform}. The behavior of all these curves resembles the Gaussian p.d.f.\ case, which has already been discussed.

If we set a finite amplitude $|\alpha|>0$ for the optical field, the eigenvalues of $\hat{M}_{\text{min}}$ are continuous.
They both start from the prior mean value $g_0$ at $\tau_c=0^+$ and show large oscillations in time.
In Fig.~\ref{costuniform} it is seen that $\bar{C}_{\text{min}}$ always starts from 
the prior variance $\sigma^2$ and reaches a 
minimum that depends on $|\alpha|$. As in the Gaussian prior p.d.f.\ case, the absolute value of $\alpha$ is sufficient to characterize completely these minima.

%%%%%%%%%%%%%%%%%%%%%%%%%%%%%%%%%%%%%%%%%%%%%%%%%%%%%%%%%%%%%%%%%%%%%%%%%%%%%%%%%%%%%%%%%%%%%%%%%%%%%%%%%%%%%%
%%%%%%%%%%%%%%%%%%%%%%%%%%%%%%%%%%%%%%%%%%%%%%%%%

%%%%%%%%%%%%%%%%%%%%%%%%%%%%%%%%%%%%%%%%%%%%%%%%%%%%%%%%%%%%%%%%%%%%%%%%%%%%%%%%%%%%%%%%%%%%%%%%%%%%

\begin{figure}
\includegraphics[width=8cm]{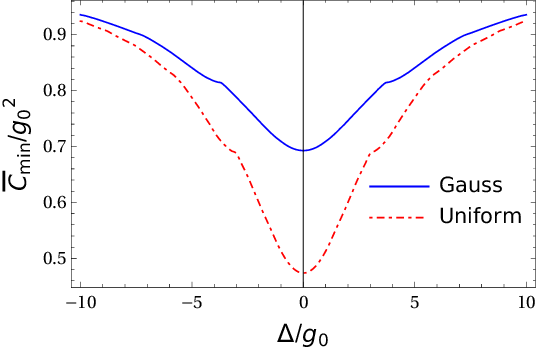}
\caption{The average minimum cost of error $\bar{C}_{\text{min}}/g_0^2$ reached at $\tau_c=\tau_c^*$ as a function of the detuning $\Delta$. The lowest average minimum cost of error is at resonance $\Delta=0$. 
We set $\gamma\tau_f=0$, $\alpha=0$ and $\sigma^2/g^2_0=1$.}
\label{costdet}
\end{figure}

{\it Role of the detuning $\Delta$ and the flight time $\tau_f$.}---In order to demonstrate the properties of non-zero detuning in a simple way we have considered set $\alpha=0$, $\gamma\tau_f=0$, and  
$\tau_c=\tau_c^*$, where the average minimum cost of error reaches its minimum. Figure~\ref{costdet}
shows that the minimum of the average minimum cost of error occurs at $\Delta=0$, for both the Gaussian and the uniform prior p.d.f. Non-zero detuning decreases the probability of the 
transition occurring in the TLS and, increasing the the average cost of error. Another interesting feature of the off-resonant case is that for $g_0 \tau_c \to \infty$, $\bar{C}_{\text{min}}$ 
does not approach $\sigma^2$ as in Figs.~\ref{fig:CostGauss} and~\ref{costuniform}, but a value depending on both $\Delta$ and the prior variance $\sigma^2$.

\begin{figure}
\includegraphics[width=8cm]{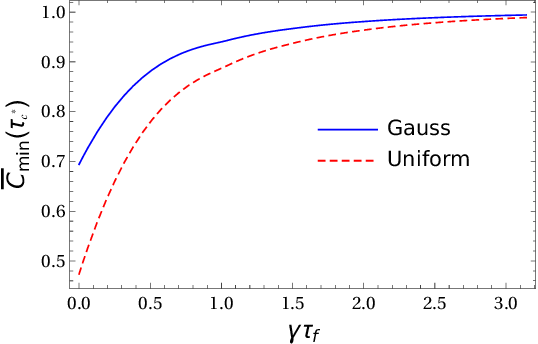}
\caption{The average minimum cost of error $\bar{C}_{\text{min}}/g_0^2$ reached at $\tau_c=\tau_c^*$ as a function of $\gamma\tau_f$. The minimum is reached when $\gamma\tau_f=0$ and 
approaches its limit value $\sigma^2/g^2_0=1$ with increasing $\gamma\tau_f$. We set $\Delta=0$ and $\alpha=0$.}
\label{costtimeflight}
\end{figure}

The influence of the flight time $\tau_f$ on the estimation scenario is clearly destructive, as we have shown in the previous subsections. Therefore, it is interesting to compare 
these deleterious effects on the two different prior p.d.f.\ considered in this work. Due to our previous findings we have set $\Delta=0$, initial single-mode field in the ground state, 
i.e., $\alpha=0$, and $\tau_c=\tau_c^*$. 
Fig.~\ref{costtimeflight} shows that the average minimum cost of error at $\tau_c^*$ reaches its minimum for $\gamma\tau_f=0$, and approaches its maximum $\sigma^2$ when $\gamma\tau_f\to\infty$.

%%%%%%%%%%%%%%%%%%%%%%%%%%%%%%%%%%%%%%%%%%%%%%%%%%%%%%%%%%%%%%%%%%%%%%%%%%%%%%%%%%%%%%%%%%%%%%%%%%%%%%%%
%%%%%%%%%%%%%%%%%%%%%%%%%%%%%%%%%%%%%%%%%%%%%%%%%%%%%%%%%%%%%%%%%%%%%%%%%%%%%%%%%%%%%%%%%%%%%%%%%%%%

%%%%%%%%%%%%%%%%%%%%%%%%%%%%%%%%%%%%%%%%%%%%%%%%%%%%%%%%%%%%%%%%%%%%%%%%%%%%%%%%%%%%%%%%%%%%%%%%%%%%%%%%
%%%%%%%%%%%%%%%%%%%%%%%%%%%%%%%%%%%%%%%%%%%%%%%%%%%%%%%%%%%%%%%%%%%%%%%%%%%%%%%%%%%%%%%%%%%%%%%%%%%%

In summary, we have been able to identify the most ideal scenario for the implementation of a MMSE estimator. Non-zero detuning, the occurrence of the spontaneous emission and 
initial states of the field with non-zero mean photon number should be avoided. If this situation is approximately achievable in some experimental setup than the interaction time $\tau_c$ 
has to be fixed to values between $0.6/g_0$ and $0.7/g_0$, which is before the appearance of the so-called collapse phenomena in the population inversion of the TLS~\cite{Wolfgang}.

\subsection{Comparison with experiments}

In this section we analyze the physical boundaries of our model proposed in Sec. \ref{I}. Here, we consider a more realistic scenario, where inside the cavity the spontaneous decay $\gamma$ of 
the TLS and the damping rate $\kappa$ of the single--mode field are present. In order to see the boundaries of our model, we take two experimental works: one in the strong coupling regime \cite{Colombe}; and the 
other one in the intermediate coupling regime \cite{Ritter}. In these experimental works the cavity mode experiences no gains from the outer world, i.e. the mean number of thermal photons is very low, and 
therefore the evolution can be effectively described by a Markovian master equation
\begin{eqnarray}\label{MEDecayDamping}
\dot{\hat{\rho}}&=&-i\left[\hat{H},\hat{\rho}\right]+\kappa\left(\hat{a}\hat{\rho} \hat{a}^\dagger-\frac{1}{2}\left\{\hat{a}^{\dagger}\hat{a},\hat{\rho}\right\}\right) \nonumber \\
&+&\gamma \left(\hat{\sigma}_-\hat{\rho} \hat{\sigma}_+-\frac{1}{2}\left\{\hat{\sigma}_+\hat{\sigma}_-,\hat{\rho}\right\}\right),
\end{eqnarray}
where the Hamiltonian $\hat{H}$ is given in Sec. \ref{I} and $\{.,.\}$ is the anticommutator. For the sake of simplicity, we consider the detuning $\Delta=0$ and a pure initial state
$\ket{e}\ket{0}$, i.e., the TLS is in the excited state and the cavity in the ground state. The state of the TLS system upon leaving the cavity is obtained from \eqref{MEDecayDamping} and yields
\begin{equation}\label{rhoatomDecayDamping}
\hat{\rho}(t)=\begin{bmatrix}
f(t) & 0\\
0 & 1-f(t)
\end{bmatrix},
\end{equation}
where
\begin{eqnarray}
f(t)& = & e^{-(\gamma+\kappa)t/2}\Big[ -8 g^2 \frac{1+ \cosh \left(\Omega t/2\right)}{\Omega^2} \nonumber \\
& &+\frac{(\gamma -\kappa)^2 \cosh \left(\Omega t/2\right)}{\Omega^2}+\frac{(\kappa -\gamma ) \sinh \left(\Omega t/2\right)}{\Omega} \Big], \nonumber
\end{eqnarray}
with $\Omega=\sqrt{(\gamma-\kappa)^2-16g^2}$.

Starting from the density matrix \eqref{rhoatomDecayDamping} we can apply the formalism of the MMSE estimator to find the average minimum cost of error in \eqref{mincost}. According to the Bayesian
formulation of the estimation problem with a quadratic cost function, those strategies and situations are more advantageous where the average cost of error is the smallest. Therefore, here we only analyze 
the average minimum cost of error with values of $\gamma$ and $\kappa$ taken from the experimental papers \cite{Colombe} and \cite{Ritter} and compare with our ideal model in Sec. \ref{I} for both a 
Gaussian and a uniform prior p.d.f., respectively.

%%%%%%%%%%%%%%%%%%%%%%%%%%%%%%%%%%%%%%%%%%%%%%%%%%%%%%%%%%%%%%%%%%%%%%%%%%%%%%%%%%%%%%%%%%%%%%%%%%%%
\begin{figure}
\centering
\begin{subfigure}{\linewidth}% width of left subfigure
\includegraphics[width=8cm]{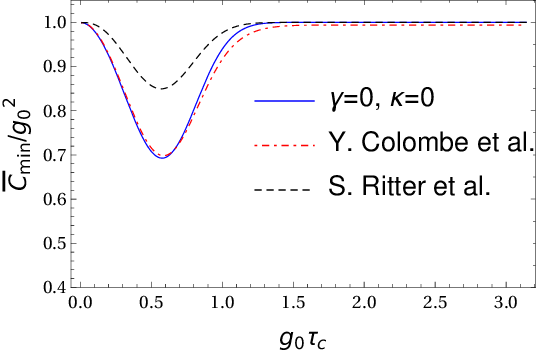}
\label{fig:expcostGauss}
\end{subfigure}
\begin{subfigure}{\linewidth}
\includegraphics[width=8cm]{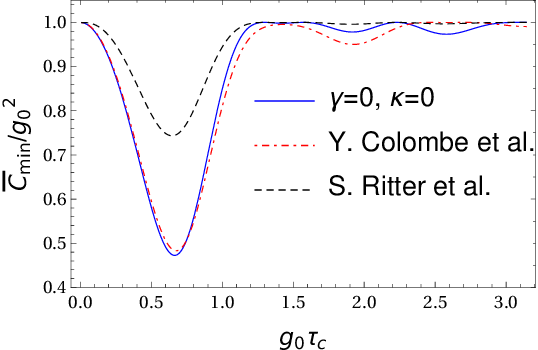}
\label{fig:expcostUniform}
\end{subfigure}
\caption{The average minimum cost of error $\bar{C}_{\text{min}}/g_0^2$ as a function of $g_0\tau_c$. The blue (solid) line represents the ideal case: $\gamma=\kappa=0$. 
The two experimental curves refer to the strong coupling regime ($\gamma/g_0=0.014$, $\kappa/g_0=0.246$) of Y. Colombe et al. \cite{Colombe} 
and to the intermediate coupling regime ($\gamma/g_0=0.6$, $\kappa/g_0=0.6$) of S. Ritter et al. \cite{Ritter}. Top: Gaussian prior p.d.f. with mean $g_0$ and variance $\sigma^2/g^2_0=1$.
Bottom: Uniform prior p.d.f with mean $g_0$ and variance $\sigma^2/g^2_0=1$. We set $\gamma\tau_f=0$, such that no spontaneous decay occurs during the time when the TLS system reaches the detectors.}
\label{expCost}
\end{figure}

Fig. \ref{expCost} shows that the average minimum cost of error is very close to the ideal model in the strong coupling regime, whereas in the intermediate coupling regime
the decoherence effects increase $\bar{C}_{\text{min}}$ for almost all interaction times, which means that the optimal estimation strategy is less informative than the ideal one. Similarly to our previous
findings, the uniform prior p.d.f. is more suitable than the Gaussian and this fact is not influenced by the addition of the decaying mechanisms inside the cavity. In summary, we have considered one of the simplest 
and demonstrative scenario, where realistic effects in experimental situations can be compared with our model. Destructive effects of decoherence sources make the estimation strategies less effective, 
as it is expected, and show that our conclusions apply only to strong coupling regimes. In the context of the MMSE estimator, this connection with the experimental parameters is more straightforward
to realize due to the more simple formalism than the one used for the ML estimators in the subsequent section. Therefore, we devote the next section only to the model of Sec. \ref{I}.

\section{Maximum-likelihood estimator}
\label{IV}

In this section we are going to determine the ML estimator. The variational problem for the average cost in Eq.~\eqref{averagecost} reads
\begin{equation}
\bar{C}[\hat{\Pi}]=\mathrm{Tr} \left \{\int_{\Theta} d\hat{\Pi}(\tilde g) z(\tilde g) \hat{\rho}(\tilde g) \right\}, \label{varML}
\end{equation}
where we are looking for those infinitesimal operators $d \hat{\Pi}(\tilde g)$ for which $\bar C$ is maximum (due to the negative sign involved in the cost function Eq.~\eqref{qcostd}). 
In order to gain insight, we employ the density matrix in Eq.~\eqref{rhoatom3}
\begin{equation}
 \hat{\rho}(g)=\begin{bmatrix}
     \cos^2(g \tau_c) e^{-\gamma \tau_f} & 0 \\
     0 & 1- \cos^2(g \tau_c)e^{-\gamma \tau_f} 
  \end{bmatrix}, \nonumber
\end{equation}
where the detuning $\Delta=0$ and the initial state of the field is in the ground state. Integrals of $d \hat{\Pi}(\tilde g)$ on compact intervals result in elements of the POVM, thus the following construction
\begin{equation}
d \hat{\Pi}(\tilde g)=\begin{bmatrix}
f_I(\tilde g)+f_z(\tilde g) & f_x(\tilde g)-if_y(\tilde g)\\
f_x(\tilde g)+if_y(\tilde g) & f_I{\tilde g}-f_z(\tilde g)
\end{bmatrix}d \tilde g, \label{ansatzPi}
\end{equation}
with $f_I, f_z, f_y,$ and $f_z$ being real functions, ensures the self-adjointness of the infinitesimal generator. We are going to employ this ansatz and solve the variational 
problem in Eq.~\eqref{varML}. Only after this step we are going to impose the constraints of the POVM in Eq.~\eqref{POVM}. In the following we reconsider the two cases of the p.d.f.\ $z(g)$ 
used in Sec.~\ref{III}.

\subsection{Gaussian probability density function}

We assume again that $g$ is characterized by its mean value $g_0$ and variance $\sigma^2$. The prior p.d.f.\ is set to be Eq.~\eqref{gaussianprior} with $\Theta=\mathbb{R}$. Then the average cost function reads
\begin{multline}
\bar C=\frac{1}{\sqrt{2\pi\sigma^2}}\int_\mathbb{R} e^{-\frac{(\tilde g-g_0)^2}{2\sigma^2}}\left[f_I(\tilde g)-f_z(\tilde g) \left(1-e^{-\gamma \tau_f}\right) \right. \\
+\left.f_z(\tilde g)\cos(2\tilde g\tau_c)e^{-\gamma \tau_f}\right] d \tilde g. \label{newvarML}
\end{multline}
We have recast the variational problem to an equivalent one where we search for the real functions $f_I$ and $f_z$ such that $\bar C$ in Eq.~\eqref{newvarML} is maximum. As $\bar C$ does 
not depend on $f_x$ and $f_y$ we set them to zero. Upon applying the transformation $\tilde g\rightarrow \sigma x+g_0$, Eq.~\eqref{newvarML} becomes
\begin{multline}
\bar C=\frac{1}{\sqrt{2\pi}}\int_\mathbb{R} dx \, e^{-x^2/2}\left[f_z(\sigma x+g_0) \left(e^{-\gamma \tau_f}-1\right) \right.\\
+\left.f_z(\sigma x+g_0)\cos(2\sigma x\tau_c+2g_0b\tau_c)e^{-\gamma \tau_f}\right.\\
\left.+f_I(\sigma x+g_0)\right].\label{funcvar}
\end{multline}
The above variational problem can be solved if we focus on square integrable functions which form the Hilbert space $\mathbb{L}^2(\mathbb{R})$ (see Ref.~\cite{Reed}). We consider now the following functions:
\begin{equation}
\Psi_n(x)=e^{-x^2/2}\frac{H_n(x)}{\sqrt{\sqrt{\pi} 2^n n!}}, \quad n=0,1,2, \dots \label{basisG}
\end{equation}
where $H_n(x)$ is the $n$\textsuperscript{th}-order Hermite polynomial with the property
\begin{equation}
 \int_\mathbb{R} H_n(x) H_m(x) e^{-x^2}dx=\sqrt{\pi} 2^n n!\delta_{nm}. \nonumber
\end{equation}
Thus, the functions in Eq.~\eqref{basisG} form an orthonormal basis in $\mathbb{L}^2(\mathbb{R})$, in which the inner product is given by the integral
\begin{equation}
 \langle f,g\rangle= \int_\mathbb{R} \overline{f(x)} g(x) dx.  \nonumber
\end{equation}
In the next step we make use of the fact that every function in the Hilbert space can expanded in the orthonormal basis. Hence,
\begin{widetext}
\begin{equation}
 f_I(\sigma x+g_0)=\sum^\infty_{n=0} \gamma^{I}_n \Psi_n(x), \quad f_z(\sigma x+g_0)=\sum^\infty_{n=0} \gamma^{z}_n \Psi_n(x), \quad \text{and} \quad \cos(2 \sigma x\tau_c+2 g_0 \tau_c)
 e^{-x^2/2}=\sum^\infty_{n=0} \gamma^{c}_n \Psi_n(x), \nonumber
\end{equation}
and, with the help of an integral formula involving Hermite polynomials~\cite{Dattoli, Galue}, we have 
\begin{eqnarray}
\gamma^{c}_n=\big \langle \cos(2 \sigma x\tau_c+2g_0\tau_c)e^{-x^2/2}, \Psi_n(x) \big\rangle=\frac{\pi^{1/4}}{\sqrt{ 2^n n!}}\times \begin{cases}
        (-1)^{n/2} (2\sigma \tau_c)^n e^{-\sigma^2\tau^2_c}\cos(2g_0\tau_c) & n\text{ is even}, \\
        (-1)^{(n+1)/2} (2\sigma\tau_c)^n e^{-\sigma^2\tau^2_c}\sin(2g_0\tau_c) & n\text{ is odd}.
      \end{cases} \nonumber \\
      \label{gammac}
\end{eqnarray}
\end{widetext}
Now, upon substituting these expansions into Eq.~\eqref{funcvar} and taking into account the properties of the orthonormal basis, we obtain
\begin{eqnarray}
 \bar C=\frac{1}{\sqrt{2\sqrt{\pi}}} \left[\gamma^{I}_0 -\gamma^{z}_0 \left(1-e^{-\gamma \tau_f}\right)+\frac{e^{-\gamma \tau_f}}{\pi^{1/4}}\sum^\infty_{n=0} \gamma^{z}_n \gamma^{c}_n \right], \nonumber \\
 \label{coeffvar}
\end{eqnarray}
where we have used the relation $e^{-x^2/2}/\pi^{1/4}=\Psi_0(x)$. We observe that $\bar{C}$ depends only on $\gamma^{I}_0$, the first coefficient in the expansion of $f_I(ax+b)$, and therefore we set 
$\gamma^{I}_n=0$ for $n \neq 0$. Thus,
\begin{equation}
 f_I(\sigma x+g_0)=\frac{\gamma^{I}_0}{\pi^{1/4}} e^{-x^2/2}, \nonumber
\end{equation}
and replacing $\sigma x+g_0$ with $x$ we have
\begin{equation}
 f_I(x)=\frac{\gamma^{I}_0}{\pi^{1/4}} e^{-\frac{(x-g_0)^2}{2\sigma^2}}. \nonumber
\end{equation}
Furthermore, $\bar C$ is maximum  with respect to $\{\gamma^{z}_n\}^\infty_{n=0}$ whenever $\gamma^{z}_n=\text{constant} \times \gamma^{c}_n$ or in other words the functions 
$\cos(2 \sigma x\tau_c+2 g_0 \tau_c)e^{-x^2/2}$ and $f_z(\sigma x+g_0)$ are parallel with respect to the inner product $\langle. \,, . \rangle$. In fact, this means that
\begin{equation}
 f_z(x)= c \times \cos(2 \tau_c x) e^{-\frac{(x-g_0)^2}{2\sigma^2}}, \quad c>0.  \nonumber
\end{equation}
We recall the following condition on the POVM
\begin{equation}
 \int_\mathbb{R} d \hat{\Pi}(x)=\hat{I}, \nonumber
\end{equation}
which, due to Eq.~\eqref{ansatzPi}, is equivalent to
\begin{eqnarray}
 \int_\mathbb{R} f_I(x) dx&=&1, \text{and} \nonumber \\
 \int_\mathbb{R} f_z(x) dx&=&\sigma\int_\mathbb{R} f_z(\sigma x+g_0) dx=0. \nonumber 
\end{eqnarray}
Then $\gamma^{I}_0=1/\sqrt{2 \sqrt{\pi} \sigma^2}$ and $\gamma^{z}_n=0$ for even $n$, the latter being due to the fact that integration of a symmetric function about the origin over the whole real line is zero
and every odd term of the orthonormal basis is such a function. We hence have
\begin{equation}
 \int_\mathbb{R} \Psi_n(x)=0, \quad n\text{ is odd}. \nonumber
\end{equation}
Thus,
\begin{eqnarray}
 f_I(x)&=&\frac{1}{\sqrt{2\pi \sigma^2}} e^{-\frac{(x-g_0)^2}{2\sigma^2}}, \text{and} \nonumber \\
 f_z(\sigma x+g_0)&=&c\times \sum_{n\text{ odd}} \gamma^{c}_n \Psi_n(x). \label{firstform}
\end{eqnarray}
There is one more requirement, namely that
\begin{equation}
 \int_{\Delta} d \hat{\Pi}(x)=\hat{\Pi}(\Delta) \nonumber
\end{equation}
is a positive semidefinite operator with a spectrum confined to the interval $[0,1]$ for every compact interval $\Delta$ in $\mathbb{R}$. This equivalent to
\begin{equation}
 0\leqslant\int_{\Delta} \left[f_I(x) \pm f_z(x)\right]dx \leqslant1, \quad \forall \Delta \in \mathbb{R}. \label{POVMc}
\end{equation}
We consider the compact interval $\Delta=[a,b]$ with arbitrary $a,b \in \mathbb{R}$ and $b>a$. Using the results of Eq.~\eqref{gammac}, we have
\begin{equation}
\sum^\infty_{n\text{ odd}} \gamma^{c}_n \Psi_n(x)= -\sin(2g_0\tau_c) \sin(2 \sigma x\tau_c)e^{-x^2/2}. \nonumber
\end{equation}
In view of the above relation,
\begin{multline}
 0\leqslant\int^b_a e^{-x^2/2} \left[\frac{1}{\sqrt{2\pi}} \pm c \times \sigma \sin(2g_0\tau_c) \sin(2 \sigma x\tau_c)\right]dx\\
\leqslant1, \nonumber
\end{multline}
where we have again employed the variable transformation $x \to \sigma x+g_0$ in Eq.~\eqref{POVMc}. In order to analyze right-hand side inequality we first make some observations. 
The area under the function $e^{-x^2/2}/\sqrt{2 \pi}$ around the origin contributes the most due to the properties of the error function $\erf(x)$ \cite{Stegun} and $\sin(2 \sigma x\tau_c)e^{-x^2/2}$ is 
an odd function. Therefore, if the following inequalities
\begin{eqnarray}
 &&0 \leqslant \int^{\frac{\pi}{2 \sigma \tau_c}}_{0} e^{-x^2/2} \left[\frac{1}{\sqrt{2\pi}} - c \times y \sin(2 \sigma x\tau_c)\right]dx, \nonumber \\
 &&\int^{\frac{\pi}{2 \sigma \tau_c}}_{0} e^{-x^2/2} \left[\frac{1}{\sqrt{2\pi}} + c \times y \sin(2 \sigma x\tau_c)\right]dx \leqslant1, \nonumber \\
 &&y=\sigma \left|\sin(2g_0\tau_c)\right|, \nonumber 
\end{eqnarray}
with $\sigma, \tau_c>0$ and $2g_0\tau_c \neq \pi +k \pi$ ($k \in \mathbb{Z}$) hold, then no matter how we choose our intervals the condition~\eqref{POVMc} is fulfilled. In the case when 
$2g_0\tau_c = \pi +k \pi$ ($k \in \mathbb{Z}$), condition~\eqref{POVMc} is automatically satisfied. Making use of the error function, we obtain
\begin{eqnarray}
 &&c\leqslant \frac{2}{\sqrt{2\pi \sigma^2} \left|\sin(2g_0\tau_c)\right| e^{-2\sigma^2\tau^2_c}} \nonumber \\
 && \qquad\times \frac{\erf\left(\frac{\pi}{2 \sqrt{2} \sigma \tau_c}\right)}{\erf\left(\frac{\pi+4 i \sigma^2 \tau^2_c}{2 \sqrt{2} \sigma \tau_c}\right)+
 \erf\left(\frac{\pi-4 i \sigma^2 \tau^2_c}{2 \sqrt{2} \sigma \tau_c}\right)}=c_1, \text{and}\nonumber \\
 &&c\leqslant \frac{2}{\sqrt{2\pi \sigma^2} \left|\sin(2g_0\tau_c)\right| e^{-2\sigma^2\tau^2_c}} \nonumber \\
 && \qquad\times \frac{2-\erf\left(\frac{\pi}{2 \sqrt{2} \sigma \tau_c}\right)}{\erf\left(\frac{\pi+4 i \sigma^2 \tau^2_c}{2 \sqrt{2} \sigma \tau_c}\right)+
 \erf\left(\frac{\pi-4 i \sigma^2 \tau^2_c}{2 \sqrt{2} \sigma \tau_c}\right)}=c_2. \label{onc12cond} 
\end{eqnarray}
As our original task was to maximize the average cost function $\bar C$, therefore, the relevant functions read
\begin{eqnarray}
 f_I(x)&=&\frac{1}{\sqrt{2\pi \sigma^2}} e^{-\frac{(x-g_0)^2}{2\sigma^2}}, \text{and}\nonumber \\
 f_z(x)&=&-c_{\max} \sin(2g_0\tau_c) \sin\big(2 \tau_c (x-g_0)\big) e^{-\frac{(x-g_0)^2}{2\sigma^2}}, \nonumber \label{lastform}
\end{eqnarray}
with $c_{\max}=\min\{c_1,c_2 \}$. Together with the ansatz~\eqref{ansatzPi} we have determined the ML estimators. Finally, the maximum of the average cost function reads
\begin{multline}
 \bar C_{\max}=\frac{1}{\sqrt{4 \pi \sigma^2}}\\
+c_{\max}\frac{e^{-\gamma \tau_f}}{\sqrt{2}} e^{-2\sigma^2\tau^2_c}\sin^2(2g_0\tau_c)\underbrace{\sum_{n\text{ odd}} \frac{(2\sigma^2\tau^2_c)^n}{n!}}_{\sinh(2\sigma^2\tau^2_c)}\\
=\frac{1}{\sqrt{4 \pi \sigma^2}}+c_{\max} e^{-\gamma \tau_f}\frac{1-e^{-4\sigma^2\tau^2_c}}{2\sqrt{2}} \sin^2(2g_0\tau_c).  \label{avcostMLGauss}
\end{multline}
The three inconclusive cases identified for the MMSE estimator, i.e., $\gamma \tau_f \gg 1$, $\tau_c=\pi/g_0$, and $\tau_c=\pi/(2g_0)$, reduce the value of $\bar C_{\max}$. 
It becomes clear that, whichever strategy is adopted, these cases should be avoided. The conditional p.d.f.\ in Eq.~\eqref{condpdf}, $p(\tilde g | g)$, is not an even function of the variable 
$g-\tilde g$, and therefore the ML estimate will be biased.

The average estimator reads
\begin{eqnarray}
E[\tilde{g}|g] & = & \int_\mathbb{R} \tilde g p(\tilde g | g) d \tilde g= g_0+4 \sqrt{5 \pi } c_{\max } \sigma^2 \tau_c \label{avestimatorMLEgauss} \\
& \times & e^{-2 \sigma^2 \tau^2_c-\gamma \tau_f} \left[e^{\gamma \tau_f}-2\cos^2(g \tau_c)\right] \sin(2g_0\tau_c). \nonumber
\end{eqnarray}
Here, the quantum Cram\'er--Rao inequality has the same form as in Eq.~\eqref{eq:lbound}, i.e.,
\begin{equation}
\mathrm{E}\left[(\tilde g -g)^2|g\right] \geqslant
\frac{|x'(g)|^2}{\mathrm{Tr}\big\{\hat{\rho}(g)\hat{L}^2 \big\}} \label{MLcr}
\end{equation}
but
\begin{equation}
 x'(g)= \int^{\infty}_{-\infty} \tilde g \mathrm{Tr}\{\frac{\partial}{\partial g} \hat{\rho}(g) d \hat{\Pi}(\tilde g)\}, \nonumber
\end{equation}
and, similarly to Sec.~\ref{III},
\begin{equation}
\hat{L}=-2 \tau_c \tan(g \tau_c) \ket{e} \bra{e}+\tau_c \frac{\sin(2 g \tau_c) e^{-\gamma \tau_f}}{1- \cos^2(g \tau_c)e^{-\gamma \tau_f}}\ket{g} \bra{g}. \nonumber
\end{equation}
Inserting Eq.~\eqref{ansatzPi} and Eq.~\eqref{avestimatorMLEgauss} into Eq.~\eqref{MLcr}, we obtain
\begin{eqnarray}
&&\mathrm{E}\left[(\tilde g -g)^2|g\right]\geqslant \left[1- \cos^2(g \tau_c)e^{-\gamma \tau_f} \right] \cos^2(g \tau_c)e^{-\gamma \tau_f} \nonumber \\
&&\times \left[ 8 \sqrt{5 \pi } \sigma^2 \tau_c c_{\max } \sin(2g_0\tau_c) \right]^2    e^{-4 \sigma^2 \tau^2_c}.
\label{MLcrbound}
\end{eqnarray}

\subsection{Uniform probability density function}

As in the previous subsection we assume that the coupling strength $g$ has mean value $g_0$ and variance $\sigma^2$. The prior p.d.f.\ is set to be Eq.~\eqref{uniformaprior} 
with $\Theta=[g_0-\sqrt{3} \sigma,g_0+\sqrt{3} \sigma ]$. Now, the average cost function reads
\begin{multline}
\bar C=\frac{1}{2\sqrt{3}\sigma} \int^{g_0+\sqrt{3} \sigma}_{g_0-\sqrt{3} \sigma} \left[f_I(\tilde g)-f_z(\tilde g) \left(1-e^{-\gamma \tau_f}\right) \right. \\
+\left.f_z(\tilde g)\cos(2\tilde g\tau_c)e^{-\gamma \tau_f}\right] d \tilde g. \label{newvar2ML}
\end{multline}
We employ the transformation $\tilde g \rightarrow \sqrt{3} \sigma x+g_0$ and obtain
\begin{multline}
\bar C=\frac{1}{2} \int^{1}_{-1} \left[f_I\left(\sqrt{3} \sigma x+g_0\right)\right.\\
\left.-f_z\left(\sqrt{3} \sigma x+g_0\right) \left(1-e^{-\gamma \tau_f}\right) \right. \\
+\left.f_z\left(\sqrt{3} \sigma x+g_0\right)\cos\left(2\sqrt{3} \sigma x \tau_c+2g_0 \tau_c\right)e^{-\gamma \tau_f}\right] dx. \nonumber
\end{multline}
This time the Hilbert space is $\mathbb{L}^2\left([-1,1]\right)$ and we choose the following orthonormal basis~\cite{Reed}:
\begin{eqnarray}
\Psi_{n,e}(x)&=&\frac{1}{\sqrt{2}} \cos(n \pi  x), \label{basisU} \\
\Psi_{n,o}(x)&=&\frac{1}{\sqrt{2}} \sin(n \pi  x), \quad n \in \mathbb{Z}. \nonumber  
\end{eqnarray}
where $\Psi_{0,e}(x)=1/\sqrt{2}$ and $\Psi_{0,o}(x)=0$. Every function can expanded in this orthonormal basis. Thus,
\begin{eqnarray}
 f_I\left(\sqrt{3} \sigma x+g_0\right)&=&\sum_{i=e,0}\sum^\infty_{n=0} \gamma^{I}_{n,i} \Psi_{n,i}(x), \nonumber \\
 f_z\left(\sqrt{3} \sigma x+g_0\right)&=&\sum_{i=e,0}\sum^\infty_{n=0} \gamma^{z}_{n,i} \Psi_{n,i}(x), \nonumber \\
 \cos\left(2\sqrt{3} \sigma x \tau_c+2g_0 \tau_c\right)&=&\sum_{i=e,0}\sum^\infty_{n=0} \gamma^{c}_{n,i} \Psi_{n,i}(x), \nonumber
\end{eqnarray}
and 
\begin{eqnarray}
 \gamma^{c}_{n,e}&=&\big \langle \cos\left(2\sqrt{3} \sigma x \tau_c+2g_0 \tau_c\right), \Psi_{n,e}(x) \big\rangle \nonumber \\
 &=&\frac{4\sqrt{3}\sigma \tau_c \sin(2\sqrt{3}\sigma \tau_c) \cos(2g_0 \tau_c)}{\sqrt{2}(12 \sigma^2 \tau^2_c-n^2 \pi^2)} \cos(n \pi), \text{and} \nonumber \\ 
 \gamma^{c}_{n,o}&=&\big \langle \cos\left(2\sqrt{3} \sigma x \tau_c+2g_0 \tau_c\right), \Psi_{n,o}(x) \big\rangle \nonumber \\
 &=&-\frac{n \pi \sin(2\sqrt{3}\sigma \tau_c) \sin(2g_0 \tau_c)}{\sqrt{2}(12 \sigma^2 \tau^2_c-n^2 \pi^2)} \cos(n \pi).  \label{gammac2}
\end{eqnarray}
Now, taking into account the properties of this orthonormal basis, we obtain
\begin{eqnarray}
 \bar C=\frac{1}{2} \left[\gamma^{I}_0 -\gamma^{z}_0 \left(1-e^{-\gamma \tau_f}\right)+e^{-\gamma \tau_f} \sum_{i=e,o} \sum^\infty_{n=0} \gamma^{z}_{n,i} \gamma^{c}_{n,i} \right], \nonumber 
\end{eqnarray}
a very similar expression to Eq.~\eqref{coeffvar}. We observe again that $\bar C$ depends only on $\gamma^{I}_0$ and therefore we set $\gamma^{I}_n=0$ for $n \neq 0$. The condition on the POVM
\begin{equation}
 \int^{g_0+\sqrt{3} \sigma}_{g_0-\sqrt{3} \sigma} d \hat{\Pi}(x)=\hat{I}, \nonumber
\end{equation}
results in
\begin{equation}
 \int^{g_0+\sqrt{3} \sigma}_{g_0-\sqrt{3} \sigma} f_I(x) dx =\int^{g_0+\sqrt{3} \sigma}_{g_0-\sqrt{3} \sigma} \frac{\gamma^{I}_0}{\sqrt{2}}dx=1 \nonumber
\end{equation}
and thus $\gamma^{I}_0=1/(\sqrt{6}\sigma)$. Similarly to the previous subsection, we have
\begin{equation}
 \int^{g_0+\sqrt{3} \sigma}_{g_0-\sqrt{3} \sigma} f_z(x) dx =\sqrt{3} \sigma \int^{1}_{-1} f_z\left(\sqrt{3} \sigma x+g_0\right) dx=0, \nonumber
\end{equation}
which yields $\gamma^{z}_{0,e}=0$. As we would like to maximize $\bar C$, therefore, we set $\gamma^{z}_{n,i}=\text{constant} \times \gamma^{c}_{n,i}$ for $n \neq 0$ ($i \in \{e,o\}$), a similar geometrical 
strategy to the one employed in the previous subsection. Thus,
\begin{eqnarray}
 f_I(x)&=&\frac{1}{2\sqrt{3} \sigma}, \text{and}\label{formalsol} \\
 f_z(x)&=&c \times\left[\cos(2x\tau_c)-\frac{\sin(2\sqrt{3}\sigma \tau_c)\cos(2 g_0 \tau_c)}{2\sqrt{3}\sigma \tau_c}\right], \nonumber
\end{eqnarray}
with $c>0$. Imposing the constraint that
\begin{equation}
 \int_{\Delta} d \hat{\Pi}(x)=\hat{\Pi}(\Delta) \nonumber
\end{equation}
is a positive semidefinite operator with a spectrum confined to the interval $[0,1]$ for every compact interval $\Delta$ in $[g_0-\sqrt{3} \sigma,g_0+\sqrt{3} \sigma ]$, we find
\begin{equation}
 0\leqslant\int^b_a \left[f_I(x) \pm f_z(x)\right]dx \leqslant1, \label{POVMc2}
\end{equation}
where $b\leqslant g_0+\sqrt{3} \sigma$ and $a\geqslant g_0-\sqrt{3} \sigma$. After performing the definite integral, we get
\begin{eqnarray}
 &&0\leqslant x \pm \frac{c}{\tau_c} \left[ \sin(x 2\sqrt{3}\sigma \tau_c)\cos(y 2 g_0 \tau_c)\right. \label{Ineq12} \\
 &&\left. -x \sin(2\sqrt{3}\sigma \tau_c)\cos(2 g_0 \tau_c) \right] \leqslant 1, \quad x=\frac{b-a}{2 \sqrt{3}\sigma}\in [0,1], \nonumber \\
 && y=\frac{b+a}{2 g_0}\in \left[1-\sqrt{3}\sigma(1-x)/g_0,1+\sqrt{3}\sigma(1-x)/g_0\right], \nonumber
\end{eqnarray}
with $g_0 \neq 0$. It is interesting to note the extreme cases $x=0$ and $x=1$, when the term
\begin{multline}
 f_{\pm} (x,y,c)=x \pm \frac{c}{\tau_c} \left[ \sin(x 2\sqrt{3}\sigma \tau_c)\cos(y 2 g_0 \tau_c)\right. \\
 \left. -x \sin(2\sqrt{3}\sigma \tau_c)\cos(2 g_0 \tau_c) \right] \label{fpm}
\end{multline}
is equal to $0$ and $1$, respectively. The functions $f_{+}(x,y,c)$ and $f_{-}(x,y,c)$ are continuous in $x$ and have extrema, where they can violate the conditions of being smaller than $1$ 
and greater than $0$. The strategy is to find these points $x_{\text{ext}}=x_{\text{ext}}(c)$. Upon replacing these back to into Eq.~\eqref{Ineq12} one is able to find $c_{\max}$. In order to 
demonstrate the procedure, let us consider $2\sqrt{3}\sigma \tau_c=2 g_0 \tau_c=\pi/2$. Then, Eq.~\eqref{fpm} reads
\begin{eqnarray}
 &&f_{\pm} (x,y,c)=x \pm \frac{c}{\tau_c} \sin\left(x\frac{\pi}{2}\right)\cos\left(y \frac{\pi}{2}\right), \nonumber \\
 && x\in [0,1], \quad y\in \left[x,2-x\right]. \nonumber
\end{eqnarray}
The two extrema of $f_{-} (x,y,c)$ are found at $y=x$ (minimum) and $y=2-x$ (maximum), yielding
\begin{eqnarray}
 1-\frac{c \pi}{2 \tau_c} \cos \left(x^{-}_{\min} \pi \right)=0, \nonumber \\
 1+\frac{c \pi}{2 \tau_c} \cos \left(x^{-}_{\max} \pi \right)=0. \nonumber
\end{eqnarray}
These equations, together with Eq.~\eqref{Ineq12}, result in
\begin{equation}
 c \leqslant \frac{2 \tau_c}{\pi}=c_1. \nonumber
\end{equation}
$f_{+} (x,y,c)$ has two extrema  at $y=x$ (maximum) and $y=2-x$ (minimum), and therefore we have
\begin{equation}
 c \leqslant \frac{2 \tau_c}{\pi}=c_2=c_1. \nonumber
\end{equation}
Finally, the task to maximize $\bar C$ yields
\begin{equation}
 c_{\max}= \min\{c_1,c_2 \}=c_1. \label{lastform2}
\end{equation}
The functions defining the ML estimator through the ansatz~\eqref{ansatzPi} finally read
\begin{eqnarray}
 f_I(x)&=&\frac{1}{2\sqrt{3} \sigma}, \text{and}\nonumber \\
 f_z(x)&=&c_{\max} \times\left[\cos(2x\tau_c)-\frac{\sin(2\sqrt{3}\sigma \tau_c)\cos(2 g_0 \tau_c)}{2\sqrt{3}\sigma \tau_c}\right]. \nonumber
\end{eqnarray}
The maximum of the average cost function is 
\begin{multline}
  \bar C_{\max}=\frac{1}{2\sqrt{3}\sigma}+c_{\max} \left[\frac{1}{2}-\frac{\sin^2(2\sqrt{3}\sigma \tau_c)\cos^2(2 g_0 \tau_c)}{12 \sigma^2 \tau^2_c} \right. \\
  +\left. \frac{\sin(4\sqrt{3}\sigma \tau_c)\cos(4 g_0 \tau_c)}{8 \sqrt{3}\sigma \tau_c}\right] e^{-\gamma \tau_f}. \label{unilast}
\end{multline}
The conditional p.d.f.\ $p(\tilde g | g)$ in Eq.~\eqref{condpdf} is again not an even function of the variable $g-\tilde g$. Therefore the ML estimate, as in the case of the prior Gaussian p.d.f., will be biased.

In the special case $2\sqrt{3}\sigma \tau_c=2 g_0 \tau_c=\pi/2$ discussed earlier,
\begin{equation}
 \bar C_{\max}=\frac{1}{4g_0} \left(2+e^{-\gamma \tau_f}\right), \nonumber
\end{equation}
and the average estimator reads
\begin{eqnarray}
E[\tilde{g}|g]&=& \int^{\frac{\pi}{2 \tau_c}}_{0} \tilde g p(\tilde g | g) d \tilde g \nonumber \\
&=&g_0+\frac{4g_0}{\pi^2}\left[1-2e^{-\gamma \tau_f}\cos^2\left(\frac{g}{g_0}\frac{\pi}{4}\right)\right]. \nonumber
\end{eqnarray}
Furthermore the inequality for the mean--squared error yields
\begin{eqnarray}
&&\mathrm{E}\left[(\tilde g -g)^2|g\right]\geqslant \frac{16 g^2_0}{\pi^4}\frac{1- \cos^2\left(\frac{\pi}{4} \frac{g}{g_0} \right)e^{-\gamma \tau_f}}{ \sin^2\left(\frac{\pi}{4} \frac{g}{g_0} \right)} \nonumber \\
&& \times \sin^2\left(\frac{\pi}{2} \frac{g}{g_0} \right) e^{-\gamma \tau_f}. 
\end{eqnarray}

\subsection{Numerical results}

%%%%%%%%%%%%%%%%%%%%%%%%%%%%%%%%%%%%%%%%%%%%%%%%%%%%%%%%%%%%%%%%%%%%%%%%%%%%%%%%%%%%%%%%%%%%%%%%%%%%
\begin{figure}[t!]
\centering
\begin{subfigure}{\linewidth}% width of left subfigure
\includegraphics[width=8cm]{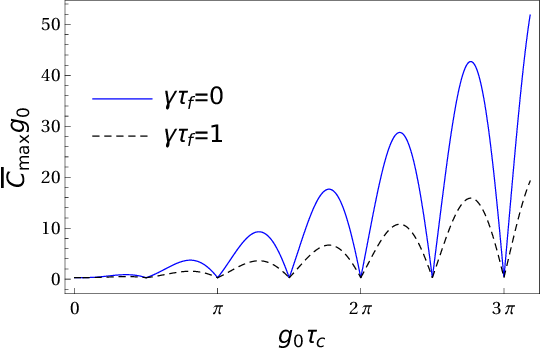}
\caption{The average maximum cost function $\bar{C}_{\max}g_0$ as a function of $g_0\tau_c$}\label{fig:MLEGCost}
\end{subfigure}
\begin{subfigure}{\linewidth}% width of left subfigure
\includegraphics[width=8cm]{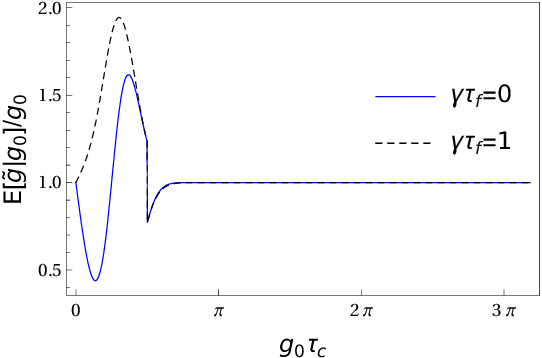}
\caption{The average estimator $E[\tilde{g}|g_0]/g_0$ as a function of $g_0\tau_c$}\label{fig:MLEGaveragestimate}
\end{subfigure}
\begin{subfigure}{\linewidth}% width of left subfigure
\includegraphics[width=8cm]{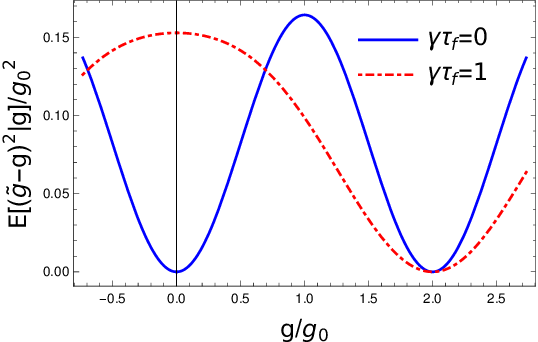}
\caption{The lower bound of the mean--squared error as a function of $g/g_0$}\label{fig:MSEML}
\end{subfigure}
\caption{Figures obtained in the case of the Gaussian prior p.d.f. We set $\sigma^2/g^2_0=1$. We see that spontaneous 
decay reduces the average cost function (a). In (b) we set $g=g_0$. We see there is a jump at $g_0\tau_c=\pi/2$ due to the properties 
of $c_{\max}$ defined in Eqs.~\eqref{onc12cond} and~\eqref{lastform}. In (c) the interaction time is $\tau_c=\pi/(4g_0)$.}
\end{figure}
%%%%%%%%%%%%%%%%%%%%%%%%%%%%%%%%%%%%%%%%%%%%%%%%%%%%%%%%%%%%%%%%%%%%%%%%%%%%%%%%%%%%%%%%%%%%%%%%%%%%

In the previous subsections we have calculated analytically the ML estimators for both the Gaussian~\eqref{gaussianprior} and the uniform~\eqref{uniformaprior} prior p.d.f. Due to the analytically 
involved solutions we have used the density matrix in Eq.~\eqref{rhoatom3}, where $\Delta=0$ and the field mode is initially in the vacuum state. Therefore, the only parameters left for the numerical 
investigations are 
the spontaneous decay rate $\gamma \tau_f$ and the interaction time $\tau_c$. We have shown in the case of the uniform prior p.d.f.\ that in Eq.~\eqref{unilast} the calculation of $c_{\max}$ is very intricate
and very much depends on the relation between the variance $\sigma^2$ and the mean $g_0$. Therefore, we consider here only the ML estimator obtained for the Gaussian prior p.d.f.

Figure~\ref{fig:MLEGCost} shows the numerical evaluation of the average maximum cost function $\bar C_{\max}$. It shows that the best time to perform the measurements is approximately 
$g_0 \tau_c=\frac{\pi}{4}+k\pi$, cf.\ Eq.~\eqref{avcostMLGauss}, with $k\in {\mathbb N}_0$ and with better results as $k$ increases. This means that longer is the interaction between the field and the TLS, 
the bigger the average cost becomes. The spontaneous decay rate $\gamma\tau_f$ affects the quality of the estimation by reducing $\bar C_{\max}$. However, on the other hand Fig.~\ref{fig:MLEGaveragestimate} 
shows that the average estimate conditioned on the mean $g_0$, a possible true value of $g$, for long interaction times is simply equal to our prior expectation. This type of dichotomy has been found by 
us~\cite{Zsolt}, where a more optimal average cost function merely leads to the reinforcement of our prior knowledge. 
Finally, the lower bound of the mean--squared error in Fig.~\ref{fig:MSEML} demonstrates the decrease of the accuracy of the estimation caused by the increase of $\gamma \tau_f$.

\section{Conclusions}
\label{V}

We have discussed Bayesian-inference approaches with a special focus on the dipole coupling of matter--field interactions. Our scheme is based on two-level systems (TLSs) which transit through a cavity and 
interact with a single-mode radiation field. The state of the TLS is subsequently measured. Spontaneous emission of the excited state of the TLS is taken into account. Our protocol assumes that all the TLS 
are prepared initially in the excited states, and that the cavity field is reset before the transit of each TLS. We have derived the 
minimum mean--square error (MMSE) estimator for both the Gaussian and the uniform probability density functions (p.d.f.)\ with given mean and variance. It has been demonstrated that the detuning between the 
TLS transition frequency and the cavity resonance frequency has a destructive effect on parameter estimation. Furthermore, spontaneous emission, as well as too long or too short interaction times, all result 
in the reinforcement or our prior expectations. In the case of resonant interactions with initial ground state of the field mode we have explicitly shown that the MMSE estimator $\hat{M}_{\text{min}}$ is 
diagonal in the basis of the qubit. Dividing $\hat{M}_{\text{min}}$ in Eq.~\eqref{eq:Moperator} by the prior mean $g_0$ results in a positive-operator valued measurement (POVM) element associated with an 
inefficient measurement scenario:
\begin{equation}
 \hat{\Pi}=\eta_1 \ket{e}\bra{e}+\eta_2 \ket{g}\bra{g}, \quad 0\leqslant \eta_1, \eta_2 \leqslant 1, \nonumber
\end{equation}
where the detection efficiencies are characterized by $\eta_1$ and $\eta_2$. These efficiencies are known functions of the priorly expected parameter values according to Eq.~\eqref{eq:Moperator}. For example, 
in the experiment described in Ref.~\cite{detect}, the final state of the TLS leaving the cavity is detected with the help of a push-out laser. This method has the potential to perform the above described 
inefficient measurement scenario. Furthermore, we have computed the average estimator and showed the biased nature of the obtained MMSE estimators. We have determined the 
lower bound of the mean--squared error with the help of a quantum Cram\'er--Rao type inequality by constructing the symmetrized logarithmic derivative of the density matrix subject to estimation. 
These calculations have been performed for initial coherent field states. The increase of the initial mean photon number decreases the effectivity of the estimation scenario due to the fact that a lot of
information is deposited in the photon number states, which in turn are traced out to obtain the state of the TLS subject to the measurements. We have also found that the 
mean--square error estimation scenario is optimal, and our prior expectations are not reinforced, when the TLS emits a photon into the single-mode field with $50\%$ probability. This is in contrast with 
many experimental situations, where every parameter is tuned such that every TLS emits a photon in the cavity thus realizing the so-called one-atom maser~\cite{maser}.

In the case of the maximum-likelihood (ML) approach the method used for the determination of the MMSE estimator cannot be applied. The observation strategy formulated with the help of the infinitesimal 
operators in Eq.~\eqref{infinitesimal} has led us to a pure mathematical problem. In general, the corresponding equations for the optimum
strategy involving the risk operator are challenging to solve~\cite{Holevo2}, but by constructing these infinitesimal operators with the help of 
square integrable functions, which form a Hilbert space with their respective inner product, we have been able to calculate the optimal POVMs. In the case of the Gaussian prior p.d.f.\ the Hilbert space is $\mathbb{L}^2(\mathbb{R})$ with the orthonormal basis 
formed by Hermite polynomials. The Hilbert space for the uniform prior p.d.f.\ case is $\mathbb{L}^2\left([-1,1]\right)$ with an orthonormal basis formed by sine and cosine functions. We have used the 
geometrical properties of these Hilbert spaces in order to optimize the average cost function. In order to be able to solve this problem we have considered the detuning to be zero and the initial state of the 
field to be the vacuum state. Aside from the main result of determining the ML estimator and the optimized average cost function, we have shown that effects of 
spontaneous emission are again destructive, and that long interaction times lead to inconclusive estimation scenarios. For both the Gaussian and the uniform a priori p.d.f., the POVM elements are diagonal 
in the basis of the qubit and as we have discussed in the case of the MMSE estimator one may implement such quantum measurements in experiment.

A few generic comments on all of the strategies presented throughout the manuscript are in order. The measurement data with the implemented POVM determines the average estimate or the a posteriori p.d.f.\ from 
which one may infer the value of the matter--field coupling constant. The lower bound of the 
mean--squared error characterizes the accuracy, but we have found that better accuracy, defined in this way, is usually associated with inconclusive scenarios. Therefore, if we would like to compare the different 
methods then it has to be done through the average cost function. In this context, we can conclude that the choice of the uniform prior p.d.f.\ is more suited for the model presented here, as shown in, e.g., 
Fig.~\ref{costdet}. In the case of a Gaussian prior p.d.f. and the MMSE estimator, we are able to compare the conclusions of Ref. \cite{Zsolt} on the estimation of the optomechanical coupling 
with those ones obtained here in this paper. It seems that this particular estimation strategy is optimal in the two different models, when the interaction time is not too long compared to one order of 
characteristic time periods of the systems. In general, this may suggest that exchange of too many excitations between the interacting systems entail a less favorable MMSE estimation scenario. A marked difference 
in the optomechanical system is the existence of a class of initial states, where the average minimum cost of error is reduced by the increase of the average excitation number of the initial state. 
This is not case for the matter-field system presented here.  

In view of recent developments in quantum information protocols based on matter--field interactions, our work can be seen as the step before the real-world application of such protocols, establishing the tools
for the optimal estimation of the dipole coupling strength. While we have not been able to solve completely all the problems related to the Bayesian approach in the context of matter--field interactions, our 
results already allow us to make several important observations, which are crucial prior to the experimental implementation of any quantum information protocol.

In closing, we note that whereas our discussion has been framed exclusively in the language of cavity QED and the interaction between TLSs and electromagnetic cavity mode, our framework may be applicable more 
broadly. For example, in hybrid optomechanical systems where a bosonic mode (corresponding to the mechanical motion of a high-quality mechanical oscillator) is coupled to a TLS, 
the dynamics is governed by a Hamiltonian similar in structure to Eq.~\eqref{HamJC}~\cite{Rabl,Ng}. The application of our techniques to this and similar scenarios is deferred to future work.

\section*{Acknowledgement}

This work is supported by the European Union's Horizon 2020 research and innovation programme under 
Grant Agreement No. 732894 (FET Proactive HOT).


\begin{thebibliography}{99}
\bibitem{commentonus} There are subtle differences between the mean--squared error, the variance of the estimator, and the variance of the mean--square error. 
An unfortunate combination of these words is used in our previous work, Ref.~\cite{Zsolt}, to describe the one and the same thing (the mean--squared error). 
In fact, the mean--squared error is the sum of the variance and the squared bias of the estimator.
\end{thebibliography}
\end{document}